\documentclass[12pt]{article}
\usepackage{hyperref}
\usepackage{amsmath}
\usepackage{graphicx}
\usepackage{amssymb} 
\usepackage{soul} 
\usepackage{xcolor}
\usepackage{framed}  
\definecolor{shadecolor}{rgb}{0.96,0.92,0.13}

\definecolor{myblue}{rgb}{0.14,0.11,0.49}
\definecolor{myred}{rgb}{0.74,0.22,0.15}
\definecolor{mygreen}{rgb}{0.05,0.52,0.42}
\definecolor{myyellow}{rgb}{0.96,0.92,0.13}
\definecolor{myorange}{rgb}{1,0.61,0.36}
\definecolor{mypurple}{rgb}{0.71,0.02,1}
\definecolor{noir}{gray}{0.} 

\newcommand{\Couleur}[1]{\textcolor{noir}{#1}}

\definecolor{htc}{rgb}{1,1,1} 

\newcommand{\abs}[1]{\left\vert#1\right\vert}
\def\be{\begin{equation}}
\def\ee{\end{equation}}
\def\bea{\begin{eqnarray}}
\def\eea{\end{eqnarray}}
\def\bc{\begin{center}}
\def\ec{\end{center}}
\def\bi{\begin{itemize}}
\def\ei{\end{itemize}}
\def\bs{\begin{slide}}
\def\es{\end{slide}}
\def\dd{\mathrm{d}}
\def\iC{\mathrm{i}}
\def\noi{\noindent}
\title{Interstellar radiation as a Maxwell field: improved numerical scheme and application to the spectral energy density}
\author{
Mayeul Arminjon$\,^a$\\
\small\it $^a$ Univ. Grenoble Alpes, CNRS, Grenoble INP
, 3SR, F-38000 Grenoble, France\\
\small\it  E-mail: Mayeul.Arminjon@3sr-grenoble.fr
}
\date{}
\begin{document}

\maketitle


\begin{abstract}
\noi The existing models of the interstellar radiation field (ISRF) do not produce a Maxwell field. Here, the recent model of the ISRF as a Maxwell field is improved by considering separately the different frequencies at the stage of the fitting. Using this improved procedure: (i) It is checked in detail that the model does predict extremely high values of the spectral energy density (SED) on the axis of a galaxy, that however decrease very rapidly when $\rho $, the distance to the axis, is increased from zero. (ii) The difference between the SED values (with $\rho =1\,$kpc or $8\,$kpc), as predicted either by this model or by a recent radiation transfer model, is reduced significantly. (iii) The slower decrease of the SED with increasing altitude $z$, as compared with the radiation transfer model, is confirmed. We also calculate the evolutions of the SED at large $\rho $. We interpret these evolutions by determining asymptotic expansions of the SED at large $z$, and also ones at large $\rho $.

\end{abstract}
\section{Introduction}

The interstellar radiation field (ISRF) in a galaxy is an electromagnetic (EM) field in a very high vacuum, hence it should be a solution of the Maxwell equations. However, the existing models for the ISRF do not take into account the full EM field with its six components coupled through the Maxwell equations. Consider, for example, the model of Chi \& Wolfendale \cite{Chi-Wolfendale1991}. It assumes an axisymmetric distribution of the volume emissivities $j_i(\lambda, \rho, z)$ of four stellar components $(i)\ (i=1,...,4)$: $j_i$  decreases exponentially with both the distance $\rho$ to the galactic axis and the altitude $z$ over the galactic central disk. The contribution of component $(i)$ to the energy density of the ISRF at some position $(\rho',z')$ and wavelength $\lambda$ is obtained by integrating $j_i(\lambda, \rho, z)g/l^2$ over the whole galactic volume. Here $l$ is the distance between the studied position and the running point in the galactic volume; $g$ describes the dust absorption and is obtained by integrating the visual extinction per unit path length over the linear path joining the studied position and the running point in the galactic volume. Other models, e.g. by Mathis, Mezger and Panagia \cite{Mathis-et-al1983}, Gordon et al. \cite{Gordon-et-al2001}, Robitaille \cite{Robitaille2011}, Popescu et al. \cite{Popescu-et-al2017}, are based on similar principles: all of these models consider quantities such as the stellar emissivity and luminosity, and the dust opacity, and they evolve the light intensity emitted by the stars by taking into account (in addition to the distance) the radiative transfer, in particular by dust absorption/ reemission. Clearly, those models do not produce an EM field, hence even less one that would be a solution of the Maxwell equations. \\

In a recent work \cite{A61}, we proposed a model applicable to the relevant ideal case of an axisymmetric galaxy, and that provides for the ISRF such an exact solution of the Maxwell equations --- a feature which, as discussed above, and to the best of our knowledge, appears to be fully new. This is indeed needed to study the relevance of a possible candidate for dark matter that emerges \cite{A57} from an alternative, scalar theory of gravity. However, it is also of astrophysical interest independently of the latter, since, as we noted, the ISRF must be an exact Maxwell field and this condition is not fulfilled by the existing models. As a step in checking the model proposed in Ref. \cite{A61}, its application to predict the variation of the spectral energy density (SED) in our Galaxy has been subjected to a first test \cite{A62}. To this purpose, the model has been adjusted by asking that the SED predicted for our local position in the Galaxy coincide with the SED determined from spatial missions by Henry, Anderson \& Fastie \cite{Henry-et-al1980}, Arendt et al. \cite{Arendt-et-al1998}, Finkbeiner et al. \cite{Finkbeiner-et-al1999}, and Porter \& Strong \cite{PorterStrong2005}. It has been found in that most recent work \cite{A62} that the spatial variation of the SED thus obtained with our model does not differ too much in magnitude from that predicted by the recent radiation transfer model of Ref. \cite{Popescu-et-al2017}, but that the SED predicted by our model: (i) is extremely high on the axis of the Galaxy --- i.e., on the axis of the axial symmetry that is assumed for the model of the Galaxy; (ii) has rather marked oscillations as function of the wavelength; and (iii) seems to decrease more slowly when the altitude $z$ increases (or rather when $\abs{z}$ increases), as compared with the radiation transfer model.\\

The aim of this paper is to present an improved numerical scheme to operate that ``Maxwell model of the ISRF", and to apply this improved scheme to check the findings (i)--(iii) above. Section \ref{SummaryModel} provides a summary of the model. Section \ref{SeparateFitting} describes the improvement of the numerical scheme. In Sect. \ref{Max_uj}, we check whether the model really predicts extremely high values of the SED on the axis of the Galaxy. Section \ref{Field_of_SED} studies the spatial variation of the SED and compares it with results of the literature. In Sect. \ref{Asymptotic}, asymptotic expansions are used to interpret the findings of the foregoing section. The Conclusion section \ref{Conclusion} is followed by Appendix \ref{Discrete_SED}, which discusses the relation between the discrete and continuous descriptions of the SED.

\section{Short presentation of the model}\label{SummaryModel}

This model has been presented in detail in Ref. \cite{A61}. 
An axisymmetric galaxy is modelled as a finite set of point-like ``stars", the azimuthal distribution of which is uniform. Those points ${\bf x}_i \ (i=1,...,i_\mathrm{max})$ are obtained by pseudo-random generation of their cylindrical coordinates $\rho ,\phi ,z$ with specific probability laws, ensuring that the distribution of $\rho $ and $z$ is approximately that valid for the star distribution in the galaxy considered, and that the set $\{{\bf x}_i \} $ is approximately invariant under azimuthal rotations of any angle $\phi $ \cite{A61}. In the present work, as in Refs. \cite{A61,A62}, $16\times 16 \times 36$ triplets $(\rho ,z,\phi )$ were thus generated, so that $i_\mathrm{max}=9216$, and the distribution of $\rho $ and $z$ is approximately that valid for the star distribution in the Milky Way.\\

The ISRF is also assumed axisymmetric, and thus depends only on $\rho $ and $z$. Since we want to describe, not the field inside the sources and in their vicinity, but instead the smoothed-out field at the intragalactic scale, we search for a solution of the source-free Maxwell equations. \hypertarget{GAZR1}{In the axisymmetric case, any time-harmonic} source-free Maxwell field is the sum of two Maxwell fields: ({\bf i}) one deriving from a vector potential having just the axial component $A_z$ non-zero, with $A_z$ obeying the standard wave equation, and ({\bf ii}) one deduced from a solution of the form ({\bf i}) by EM duality \cite{A60}. We consider for simplicity a model ISRF that has a finite frequency spectrum $(\omega_j)_{j=1,...,N_\omega }$, hence we may apply the foregoing result to each among its time-harmonic components $(j)$, and then just sum these components. Moreover, we envisage the ISRF as being indeed an EM {\it radiation} field, thus excluding from consideration the purely magnetic part of the interstellar EM field \cite{BeckWielebinski2013}. Hence the ISRF is made of ``totally propagating" EM waves, i.e., ones without any ``evanescent" component \cite{A61,GAZR2014}. Specifically, we assume that the two scalar potentials $A_{j\,z}$ and $A'_{j\,z}$ that define the decomposition ({\bf i})-({\bf ii}) of each time-harmonic component $(j)$, mentioned above, are themselves totally propagating. In that case, both $A_{j\,z}$ and $A'_{j\,z}$ have the explicit form \cite{ZR_et_al2008,GAZR2014}:
\be\label{psi_monochrom_j}
\Couleur{\psi _{\omega_j\ S_j} \,(t,\rho,z) = e^{-\iC \omega_j t} \int _{-K_j} ^{+K_j}\ J_0\left(\rho \sqrt{K_j^2-k^2}\right )\ e^{\iC k \, z} \,S_j(k)\, \dd k},
\ee
with $\omega_j$ the angular frequency, $K_j:=\omega_j /c$, $J_0$ the first-kind Bessel function of order $0$, and where $S_j$ is some (generally complex) function of $k \in [-K_j,+K_j]$. For a totally propagating, axisymmetric EM field, but otherwise fully general, the two potentials $A_{j\,z}$ and $A'_{j\,z}$ may be different, i.e., may correspond with different ``spectra" in Eq. (\ref{psi_monochrom_j}), say $S_j$ and $S'_j$ \cite{A60}.\\

To determine these potentials, that is, to determine the spectrum functions $S_j$, we use a sum of potentials emitted by the ``stars". We assume that every ``star", each at some point ${\bf x}_i$, contributes to the global potential $A_{j\,z}$ of a given frequency $\omega _j$ ($j=1,...,N_\omega $) by a spherically symmetric scalar wave of the same frequency $\omega _j$, whose emission center is its spatial position ${\bf x}_i$ --- in order that all the directions starting from the star be equivalent. Thus, consider time-harmonic spherically symmetric solutions of the wave equation that have a given angular frequency $\omega $. It is easy to check by direct calculation that they can be either an outgoing wave, an ingoing wave, or the sum of an ingoing wave and an outgoing one, and that, up to an amplitude factor, the following is the only outgoing wave:
\be\label{psi_spher}
\psi _\omega \ (t,{\bf x}) = \frac{e^{\iC (Kr-\omega t)}} {Kr}, \qquad K:=\frac{\omega }{c},\qquad r:=\abs{{\bf x}}. 
\ee
Clearly, only that outgoing solution is relevant here, given that the point-like ``stars" must be indeed {\it sources} of radiation.
\ \footnote{\ 
The outgoing wave (\ref{psi_spher}) can also be characterized \cite{WikiSommerfeldRadCond} as being the only time-harmonic spherically symmetric solution of the wave equation that satisfies the Sommerfeld radiation condition \cite{Sommerfeld1912}. However, the Sommerfeld condition aims precisely at selecting a boundary condition in order to find only ``physical", i.e., outgoing solutions for the Helmholtz equation. The latter equation applies to general time-harmonic solutions of the wave equation. In the spherically symmetric case, the time-harmonic solutions are easy to find and the outgoing solutions are immediate to recognize.
}
Thus, the contributions of the $i$-th star to the potentials $A_{j\,z}$ and $A'_{j\,z}$ can differ only in amplitude, since both must be a multiple of 
\be\label{psi_spher_i} 
\psi _{{\bf x}_i\,\omega_j} \ (t,{\bf x}) := \psi _{\omega_j} \ (t,{\bf x}-{\bf x}_i) = \frac{e^{\iC (K_j r_i-\omega_j t)}} {K_j r_i},  
\ee
where $\ K_j:=\frac{\omega_j }{c}, \qquad r_i:=\abs{{\bf x}-{\bf x}_i}$. But there is no apparent physical reason to affect different amplitudes to the contribution of the $i$-th star to $A_{j\,z}$ and to $A'_{j\,z}$, hence we assume both of them to be equal to $\psi _{{\bf x}_i\,\omega_j}$. To determine the global potentials $A_{j\,z}$ and $A'_{j\,z}$ $(j=1,...,N_\omega )$, that generate the axisymmetric model ISRF with a finite frequency spectrum $(\omega _j)$, the sum of the spherical potentials (\ref{psi_spher_i}) emanating from the point stars is fitted to the form (\ref{psi_monochrom_j}). As noted in Ref. \cite{A61}, this is not assuming that the ISRF is indeed the sum of the radiation fields emitted by the different stars (which is not correct, due to the radiation transfers) --- because ({\it a}) the equalities (\ref{Psi-simeq-Psi'}), (\ref{Psi-simeq-Psi'-j-by-j}) or (\ref{Psi-simeq-Psi'-j-by-j-space}) below are not exact equalities but ones in the sense of the least squares, and ({\it b}) nothing is really assumed regarding the EM field of the ``star" itself, in particular we actually do not need to assume that it has the form \hyperlink{GAZR1}{({\bf i})-({\bf ii})} above (e.g. the one corresponding with two equal potentials $A_{i\,j\,z}=  A'_{i\,j\,z} = \psi _{{\bf x}_i\,\omega_j}$). \\ 

In the previous works \cite{A61,A62}, this fitting was done for all frequencies at once. That is, the following least-squares problem was considered:
\be\label{Psi-simeq-Psi'}
\sum_{j=1}^{N_\omega } \sum_{i=1} ^{i_\mathrm{max}} w_j \psi _{{\bf x}_i\,\omega_j} \cong  \sum _{j=1}^{N_\omega }  \,\psi _{\omega_j\ S_j}\qquad \mathrm{on}\ G,
\ee
where the sign $\cong$ indicates that the equality is in the sense of the least squares (the arguments of the functions varying on some spatio-temporal grid $G$), and where the numbers $w_j>0$ are the weights affected to the different frequencies. In view of the axial symmetry, the spatial position ${\bf x}$ is restricted to the plane $\phi =0$, so ${\bf x} = {\bf x}(\rho ,z)$ and 
\be\label{Grid}
G = \{(t_l,\rho _m, z_p),\ 1\leq l\leq N_t,\ 1\leq m\leq N_\rho , \ 1\leq p \leq N_z \}. 
\ee
Since the contributions of the $i$-th star to $A_{j\,z}$ and to $A'_{j\,z}$ have both been assumed to be equal to $\psi _{{\bf x}_i\,\omega_j}$, there is no possibility to distinguish between $A_{j\,z}$ and $A'_{j\,z}$, either --- whence $\psi _{\omega_j\ S_j}=A_{j\,z}=A'_{j\,z}$ on the r.h.s. of (\ref{Psi-simeq-Psi'}). The unknowns of the problem are the spectrum functions $S_j ,\quad j=1,..., N_\omega$. We determine $S_j$ by the (generally complex) values 
\be\label{S_nj}
S_{n j} := S_j(k_{n j})\quad (n=0,...,N), 
\ee
where 
\be\label{k_nj}
k_{n j} = -K_j + n\delta _j \quad (n=0,...,N),
\ee
with $\delta _j : = 2K_j/N$, is a regular discretization of the interval $[-K_j,+K_j]$ for $k$ in the integral (\ref{psi_monochrom_j}). Calculating those integrals with the ``Simpson $\frac{3}{8}$ composite rule", (\ref{Psi-simeq-Psi'}) becomes the computable least-squares problem 
\be\label{Psi-simeq-Psi'-discrete}
\sum_{j=1}^{N_\omega } \sum_{i=1} ^{i_\mathrm{max}} w_j \psi _{{\bf x}_i\,\omega_j} \cong \sum _{j = 1} ^{N_\omega } \sum _{n=0} ^N f_{n j} \,S_{n j}\qquad \mathrm{on}\ G,
\ee
with 
\be\label{f_nj}
f_{n j}(t,\rho ,z) =  a_{n j} \,J_0\left(\rho \sqrt{K_j^2 - k_{n j}^2} \right) \exp \left[ \iC \left( k_{n j} z -\omega _j\, t\right ) \right ].
\ee 
The $S_{n j}$ 's are the solved-for parameters in the least-squares problem (\ref{Psi-simeq-Psi'-discrete}). In Eq. (\ref{Psi-simeq-Psi'-discrete}), $N$ must be a multiple of $3$, and in Eq. (\ref{f_nj}) we have
\bea\label{a_nj}
a_{n j} & = & (3/8)\,\delta _j  \qquad \quad (n = 0\ \mathrm{or}\ n = N),\\
a_{n j} & = & 2\times (3/8)\,\delta _j  \quad \ (\mathrm{mod}(n,3)=0 \ \mathrm{and}\ n \ne 0\ \mathrm{and}\ n \ne N),\\
a_{n j} & = &  3\times (3/8)\,\delta _j  \quad \ \mathrm{otherwise}.
\eea
Part ({\bf i}) of the \hyperlink{GAZR1}{decomposition of the model ISRF}  then obtains as follows \cite{A61}: 
\be\label{Ephi-Brho-Bz=0}
E_\phi = B_\rho = B_z=0,
\ee
\be\label{Bphi'}
B_\phi (t,\rho ,z) = \sum _{n=0} ^N \sum _{j = 1} ^{N_\omega } R_n \,J_1\left(\rho \frac{\omega _j}{\omega _0} R_n \right) \,{\mathcal Re} \left[ F_{n j}(t,z)\right ] + O\left(\frac{1}{N^4}\right), 
\ee
\be\label{Erho'}
E_\rho (t,\rho ,z) = \sum _{n=0} ^N \sum _{j = 1} ^{N_\omega } \frac{c^2}{\omega_0 } k_n R_n\,J_1\left(\rho \frac{\omega _j}{\omega _0} R_n \right) \,{\mathcal Re} \left[F_{n j}(t,z)\right ] + O\left(\frac{1}{N^4}\right) ,
\ee
\be\label{Ez'}
E_z (t,\rho ,z) = \sum _{n=0} ^N \sum _{j = 1} ^{N_\omega } \left(\frac{c^2}{\omega_0 } k_n^2 -\omega _0 \right ) \,J_0\left(\rho \frac{\omega _j}{\omega _0} R_n \right) \,{\mathcal Im} \left[F_{n j}(t,z) \right] + O\left(\frac{1}{N^4}\right),
\ee
with \ $R_n = \sqrt{K_0^2 - k_n^2}$ \ and
\be\label{F_nj}
F_{n j}(t,z) = \left(\frac{\omega _j}{\omega _0}\right)^2\, a_n \exp \left[ \iC \left( \frac{\omega _j}{\omega _0} k_n z -\omega _j\, t\right ) \right ]\,S_{n j}.
\ee
(Here $k_n$ and $a_n$ ($0 \leq n \leq N$) are as $k_{n j}$ and $a_{n j}$ in Eqs. (\ref{k_nj}) and (\ref{a_nj}), replacing $K_j$ by $K_0 =\frac{\omega _0}{c}$, with $\omega _0$ some (arbitrary) reference frequency.) Since we assume $A_{j\,z}=A'_{j\,z}$ for the global potentials generating the model ISRF, part ({\bf ii}) of \hyperlink{GAZR1}{its decomposition} is deduced from the first part by the EM duality: 
\be\label{dual}
{\bf E}' = c{\bf B}, \quad {\bf B}' = -{\bf E}/c.
\ee
It follows from this and from (\ref{Ephi-Brho-Bz=0}) that the model ISRF, sum of these two parts, has the components (\ref{Bphi'})--(\ref{Ez'}), and that the other components are just
\be\label{othercomponents}
E_\phi = cB_\phi  ,\qquad B_\rho =-E_\rho /c,\qquad B_z = - E_z/c.
\ee

\vspace{2mm}

\section{Frequency-by-frequency fitting of the potentials}\label{SeparateFitting}

Equation (\ref{Psi-simeq-Psi'}) may be split into the different frequencies (marked by the index $j$), simply by removing the sum on $j$ from both sides of either equation. The same is true for Eq. (\ref{Psi-simeq-Psi'-discrete}). Naturally, also the weight $w_j$ may then be removed from the l.h.s., by entering the inverse $1/w_j$ into the unknown spectrum function $S_j$ on the r.h.s. Equation (\ref{Psi-simeq-Psi'-discrete}) thus becomes
\be\label{Psi-simeq-Psi'-j-by-j}
\sum_{i=1} ^{i_\mathrm{max}} \psi _{{\bf x}_i\,\omega_j} \cong \sum _{n=0} ^N f_{n j} \,S_{n j}\qquad \mathrm{on}\ G\qquad (j=1,...,N_\omega ).
\ee
At this point, one notes that both $\psi _{{\bf x}_i\,\omega_j} $ [Eq. (\ref{psi_spher_i})] and $f_{n j}$ [Eq. (\ref{f_nj})] have the same dependence on time, $\exp(-\iC\omega _j t)$, which we can hence remove also, to obtain a least-squares problem with merely the spatial variables $\rho $ and $z$:
\be\label{Psi-simeq-Psi'-j-by-j-space}
\sum_{i=1} ^{i_\mathrm{max}} \frac{e^{\iC K_j r_i}} {K_j r_i} \cong \sum _{n=0} ^N g_{n j} \,S_{n j}\qquad \mathrm{on}\ G'\qquad (j=1,...,N_\omega ),
\ee
where $G' = \{(\rho _m, z_p),\ 1\leq m\leq N_\rho , \ 1\leq p \leq N_z \}$ is the spatial grid, and
\be\label{g_nj}
g_{n j}(\rho ,z) = a_{n j} \,J_0\left(\rho \sqrt{K_j^2 - k_{n j}^2} \right) \exp \left( \iC k_{n j} z \right ).
\ee
The separation, into the different frequencies, of the fitting of the sum of the potentials emitted by the ``stars", is consistent with the linearity of the wave equation and the Maxwell equations. Moreover, the elimination of the time variable from the fitting represents an appreciable gain in computing time. We recall that, for the EM field in a galaxy, the arguments of the Bessel function $J_0$ and the angular exponential, e.g. in Eq. (\ref{g_nj}), have the huge magnitude $\abs{{\bf x}}/\lambda \sim 10^{25}$, which enforces us to use a precision better than quadruple precision in the computer programs, thus leading to slow calculations \cite{A61}. Note that the ``separate fitting", i.e. the least-squares problem (\ref{Psi-simeq-Psi'-j-by-j-space}), is not exactly equivalent to the ``grouped fitting", i.e. the least-squares problem (\ref{Psi-simeq-Psi'-discrete}) (this will be confirmed by the numerical results below): the two are slightly different ways of adjusting the global potentials (\ref{psi_monochrom_j}).
\footnote{\
If Eq. (\ref{Psi-simeq-Psi'-j-by-j}), or equivalently Eq. (\ref{Psi-simeq-Psi'-j-by-j-space}), were an exact equality instead of being an equality in the sense of the least squares, then of course it would imply Eq. (\ref{Psi-simeq-Psi'-discrete}) (with $w_j\equiv 1$) as an exact equality.
}
However, equations (\ref{Ephi-Brho-Bz=0})--(\ref{othercomponents}) apply with the separate fitting as well --- although the relevant values $S_{n j}$ are different. The separate fitting is more appropriate, because solutions corresponding with different frequencies behave independently in the Maxwell equations, and each frequency can be treated with more precision by considering it alone. Indeed a very important point is that, by switching to the separate fitting, we improve the situation regarding the ``overfitting", i.e., we decrease the ratio $R$ of the number of parameters $N_\mathrm{para} $ to the number of data $N_\mathrm{data} $: now, for each value of the frequency index $j$, we have to solve the least-squares problem (\ref{Psi-simeq-Psi'-j-by-j-space}), with $N_\mathrm{para} =N+1$ unknown parameters and $N_\mathrm{data}=N_\rho \times N_z $ data (the ``data" are the values of the l.h.s. of (\ref{Psi-simeq-Psi'-j-by-j-space}) on the spatial grid $G'$). Whereas, with the formerly used grouped fitting, we had to solve just one least-squares problem (\ref{Psi-simeq-Psi'-discrete}) with $N_\mathrm{para} =(N+1)\times N_\omega $ unknown parameters and $N_\mathrm{data}=N_t\times N_\rho \times N_z $ data. \\

On the other hand, through the processes of radiative transfer, there are indeed transfers of radiation intensity from some frequency domains to other ones, e.g. the interaction with dust leads to a transfer from higher to lower frequencies (see e.g. Fig. 3 in Ref. \cite{Gordon-et-al2001}). But these processes are not directly taken into account by the present model: not any more with the grouped fitting than with the separate fitting. They are indirectly taken into account through the adjustment of the energy density \cite{A62}, which we briefly recall now. \\

The time-averaged volumic energy density of an EM field having a finite set of frequencies, $(\omega _j)_{j=1,...,N_\omega }$, is given by \cite{A62}
\be\label{Udiscrete}
\overline{U}({\bf x}) :=\overline{\frac{\delta W}{\delta V}}({\bf x}) = \sum _{j=1} ^{N_\omega } u_j({\bf x}), \qquad u_j({\bf x}):= \frac{1}{4} \sum _{q=1} ^6 \alpha _q \abs{C^{(q)}_j({\bf x})}^2,
\ee
where the complex numbers $C^{(q)}_j({\bf x})\ (q=1,...,6)$ are the coefficients in the expansion, in time-harmonic functions, of each among the six components of the EM field:
\be\label{F(t)}
F^{(q)}(t,{\bf x}) = {\mathcal Re} \left ( \sum _{j=1} ^{N_\omega } C^{(q)}_j({\bf x}) e^{-\iC \omega _j t} \right )\qquad (q=1,...,6);
\ee
and where $\alpha _q= \epsilon _0$ for an electric field component, whereas $\alpha _q= \epsilon _0 c^2$ for a magnetic field component (here $\epsilon _0 $ is the vacuum permittivity, with $\epsilon _0 = 1/(4\pi \times 9\times 10^9)$ in SI units). For an axisymmetric EM field, it is enough to consider the plane $\phi =0$, thus ${\bf x}={\bf x}(\rho ,z)$, and we have
\be\label{u_j_rho_z}
C^{(q)}_j=C^{(q)}_j(\rho ,z),\qquad u_j=u_j(\rho ,z). 
\ee
Using in that case the decomposition \hyperlink{GAZR1}{({\bf i})-({\bf ii})}, the expressions of three among the $C^{(q)}_j$ coefficients follow directly from the expressions (\ref{Bphi'})--(\ref{Ez'}) of the corresponding components of the EM field \cite{A62}. Moreover, in the special subcase (\ref{dual}) considered here, the other components are given by (\ref{othercomponents}), whence in the same way the three remaining $C^{(q)}_j$ coefficients. \\

Now note that, in the least-squares problem (\ref{Psi-simeq-Psi'-j-by-j-space}), that we use to determine the values $S_{n j}$ allowing to compute the EM field (\ref{Bphi'})--(\ref{Ez'}) and (\ref{othercomponents}), no data relative to the intensity of the fields emitted by the point-like ``stars" has been used until now. Hence, we may multiply the l.h.s. of (\ref{Psi-simeq-Psi'-j-by-j-space}) by some number $\xi _j>0$, thus obtaining now new values $S'_{n j} =\xi _j S_{n j}\ (n=0,...,N)$ as the solution of (\ref{Psi-simeq-Psi'-j-by-j-space}). 
\footnote{\
We might even affect different weights $\xi _{i j}$ to the radiations of frequency $\omega _j$ emitted by the different stars $(i)$, in order to account for different luminosities. However, given that our aim is to determine the spectra $S_j$, each of which characterizes according to Eq. (\ref{psi_monochrom_j}) the axisymmetric radiation of frequency $\omega _j$ in the galaxy, we feel that this would not likely change the results very significantly.
}
Therefore, to adjust the model, we determine the numbers $\xi _j\ (j=1,...,N_\omega )$ so that the values $u_j({\bf x}_\mathrm{loc})$ of the SED for our local position ${\bf x}_\mathrm{loc}$ in the Galaxy and for the frequencies $\omega _j$, as given by Eq. (\ref{Udiscrete}), coincide with the measured values, as determined from spatial missions. We take the measured local values $f_{\bf x_\mathrm{loc}}(\lambda _j)$ as plotted in Ref. \cite{PorterStrong2005} (see  Appendix \ref{Discrete_SED}), and we take $\rho_\mathrm{loc} = 8$ kpc and $z_\mathrm{loc} = 0.02$ kpc, see e.g. Ref. \cite{Majaess-et-al2009}. The model thus adjusted then allows us to make predictions: in particular, predictions of the spatial variation of the SED in the Galaxy. Such predictions may then be compared with predictions of the mainstream models of the ISRF, which models are very different from the present model.

\section{Results: maximum energy density}\label{Max_uj}

In the foregoing work \cite{A62}, the same adjustment just described was used in the framework of the ``grouped fitting" (i.e. the least-squares problem 
(\ref{Psi-simeq-Psi'-discrete})). A surprising result was that found for the values of the maximum of the energy density $u_j({\bf x})$ in the Galaxy --- thus, owing to the axial symmetry (\ref{u_j_rho_z}), for the values of 
\be
u_{j \mathrm{max}} = \mathrm{Max}\{u_j(\rho _m, z_p); \ m=1,...,N_\rho ,\  p=1,...,N_z\},
\ee
found for the different spatial grids investigated, all having $\rho $ varying regularly from $\rho _0=0$ to $\rho_\mathrm{max}\simeq 10$\,kpc and $z$ varying regularly from $z_0=0$ or $z_0=-z_\mathrm{max}$ to $z_\mathrm{max}\le 1\,$kpc.
\footnote{\ \label{def_grid}
Precisely: $\rho _0:=\rho _{m=1}$\,; $\rho_\mathrm{max}:=\rho _{m=N_\rho }$ with, in this subsection, $\rho_\mathrm{max}= 10\,\mathrm{kpc}\times \frac{N_\rho -1}{N_\rho}$\,; $z_0:=z_{p=1}$ and $z_\mathrm{max}:=z_{p=N_z}$ with, in this paper, $z_\mathrm{max}=1\,$kpc and $z_0=-z_\mathrm{max}$.
}
These maximum values, which are always found at $\rho =0$, thus on the axis of symmetry, are extremely high for lower wavelengths $\lambda_j$, with $u_{j \mathrm{max}} \simeq 10^{27}\,\mathrm{eV/cm}^3$. Moreover, the value of $u_j(\rho =0,z)$ depends little on $z$ in the domain investigated.
\begin{figure}[ht]
\centerline{\includegraphics[height=6cm]{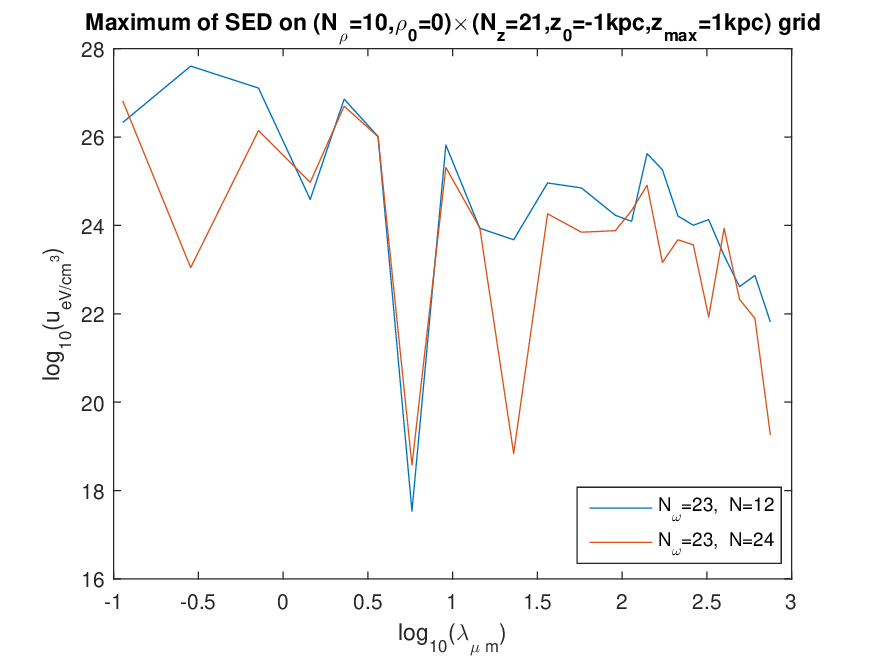}}
\centerline{\includegraphics[height=6cm]{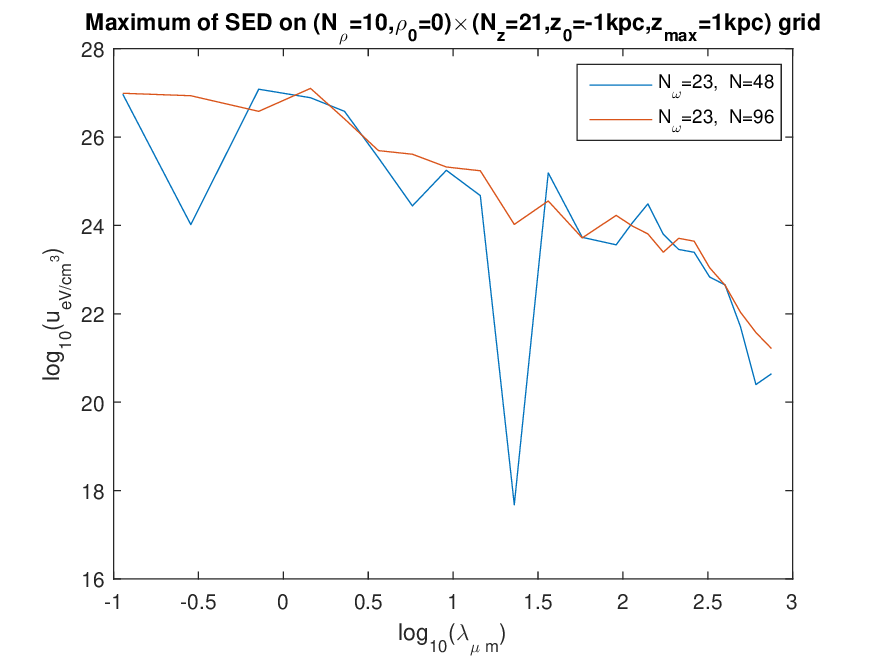}}
\centerline{\includegraphics[height=6cm]{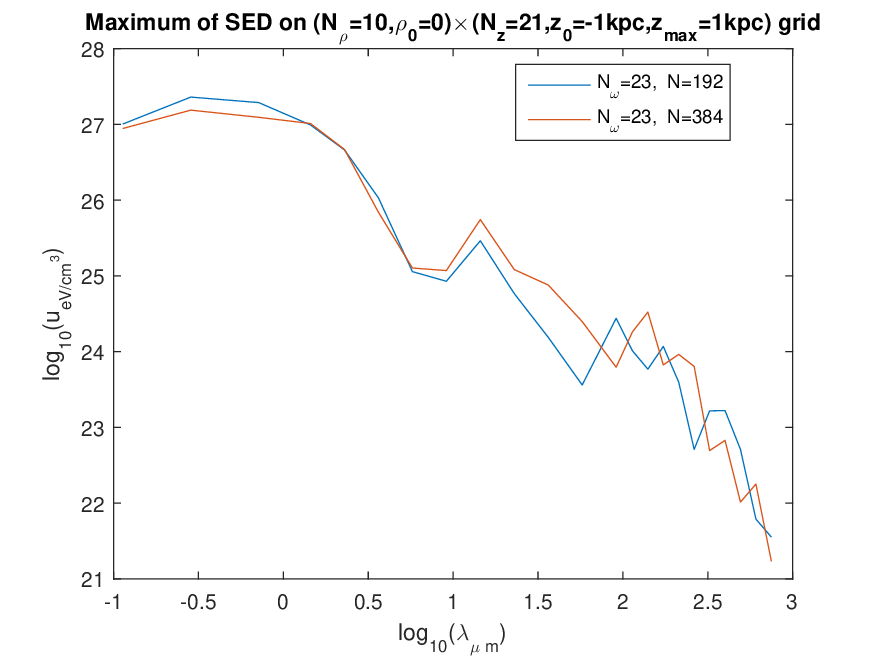}}
\caption{Effect of discretization number $N$ for: $N_\omega =23$, rough grid}
\label{23-rough}
\end{figure}
These surprisingly high values occur in a larger or smaller range of wavelengths, depending on the settings of the calculation. Therefore, the question arises whether these extremely high values are a physical effect or a numerical artefact. However, the dependence on the settings is governed by the ``amount of overfitting": less overfitting increases the range of the high values \cite{A62}. This makes it plausible that the high values might be a true prediction of the model. We will now try to check whether this is indeed the case. 

\subsection{Robustness of the high values on the axis}\label{Robust high}

In the present work based on the separate fitting (which, we argued, is more appropriate), we investigated rather systematically the question which we just asked. Since the influence of the spatial grid was found weak in the foregoing work \cite{A62}, only two grids were tried: an $(N_\rho = 10,\rho_0=0) \times (N_z = 21, z_0=-1\,\mathrm{kpc}, z_{\mathrm{max}}=1\,\mathrm{kpc})$ grid (hereafter ``rough grid"), and an $(N_\rho=20,\rho_0=0)\times (N_z=23,z_0=-1\,\mathrm{kpc}, z _{\mathrm{max}}=1\,\mathrm{kpc})$ grid (hereafter ``fine grid"). However, we investigated the influence of the fineness of the frequency mesh ($N_\omega $) and the influence of the discretization number $N$ quite in detail. [That integer $N$ is used to compute the integrals over the wavenumber $k$, e.g. the integral (\ref{psi_monochrom_j}) approximated to $\sum _{n=0} ^N f_{n j} \,S_{n j}$, see Eq. (\ref{Psi-simeq-Psi'-discrete}).] The effect of choosing $N_\omega = 23$, $N_\omega=46$, or $N_\omega=76$, was studied simultaneously with the effect of choosing $N=12$, or $N=24,48,96,192,384$, and this was done for the two different grids. \\

\begin{figure}[ht]
\centerline{\includegraphics[height=6cm]{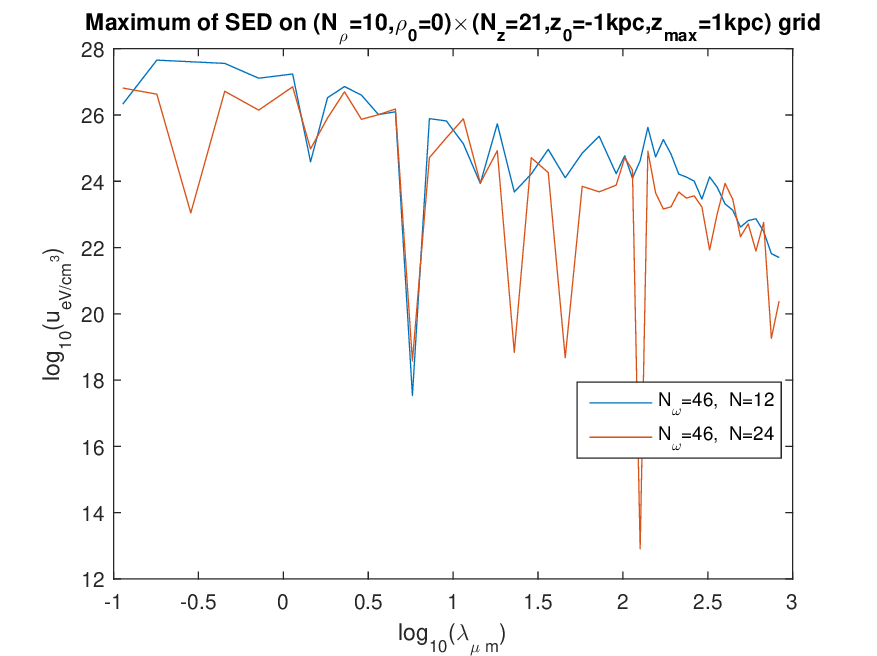}}
\centerline{\includegraphics[height=6cm]{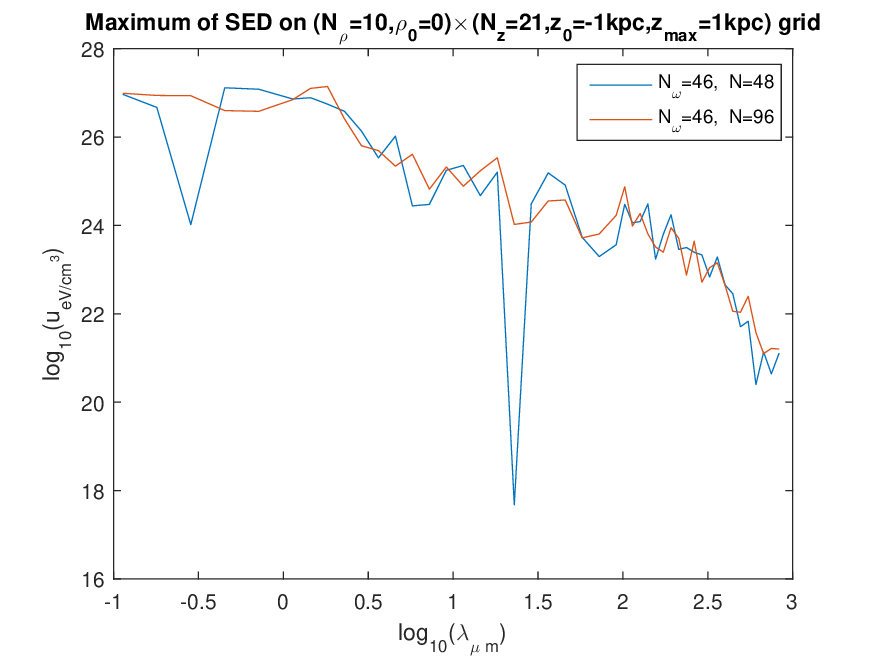}}
\centerline{\includegraphics[height=6cm]{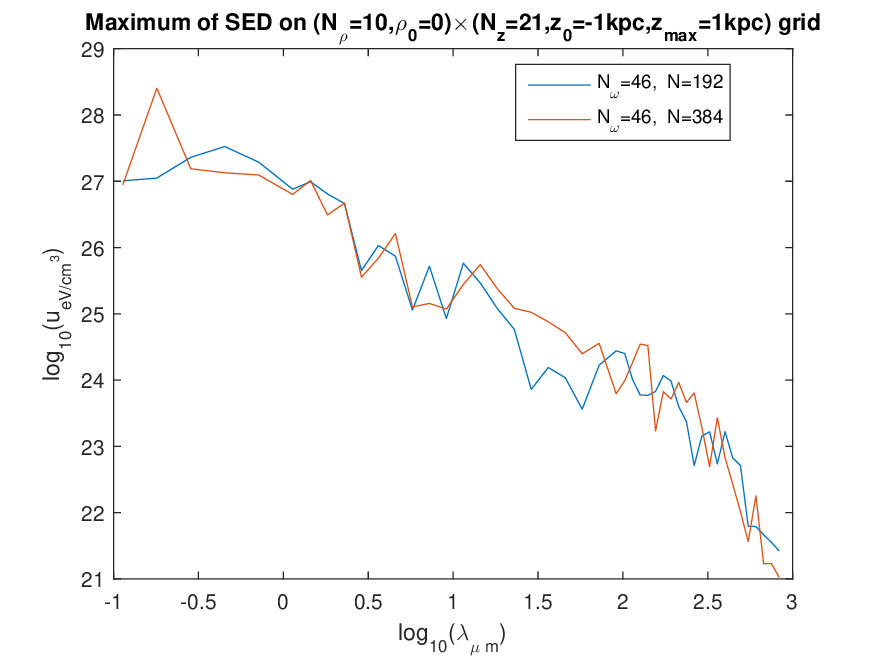}}
\caption{Effect of discretization number $N$ for: $N_\omega =46$, rough grid}
\label{46-rough}
\end{figure}

\begin{figure}[ht]
\centerline{\includegraphics[height=6cm]{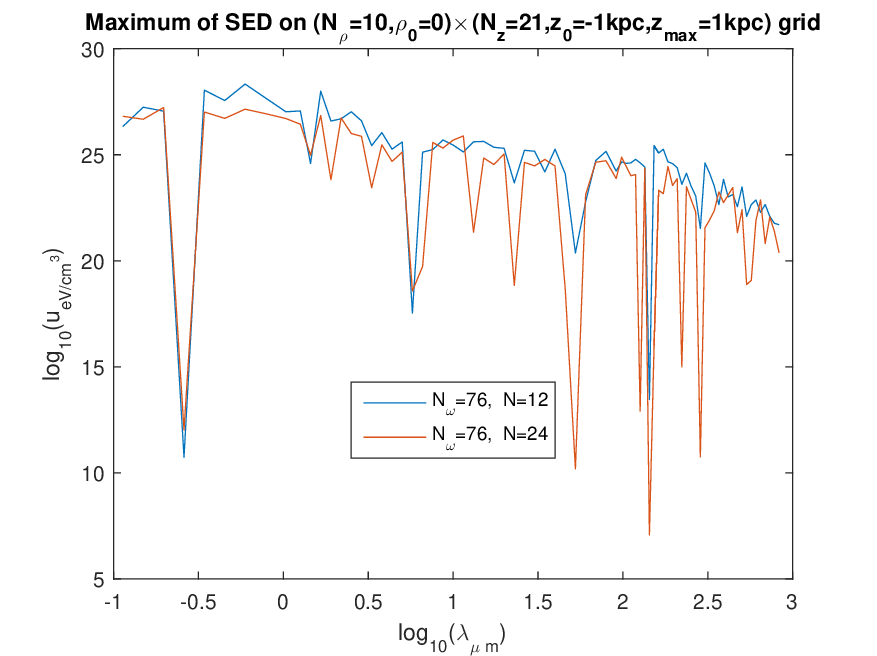}}
\centerline{\includegraphics[height=6cm]{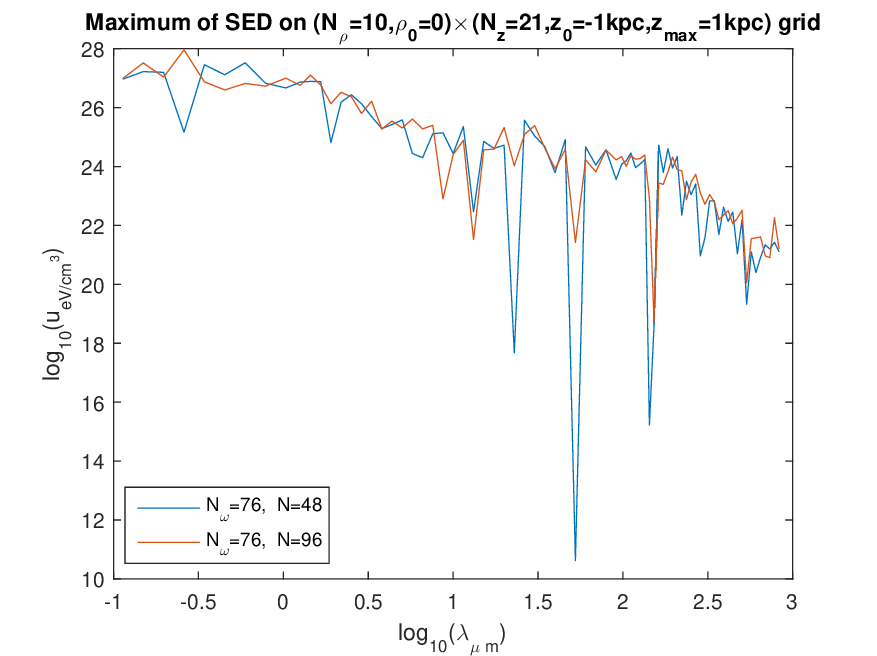}}
\centerline{\includegraphics[height=6cm]{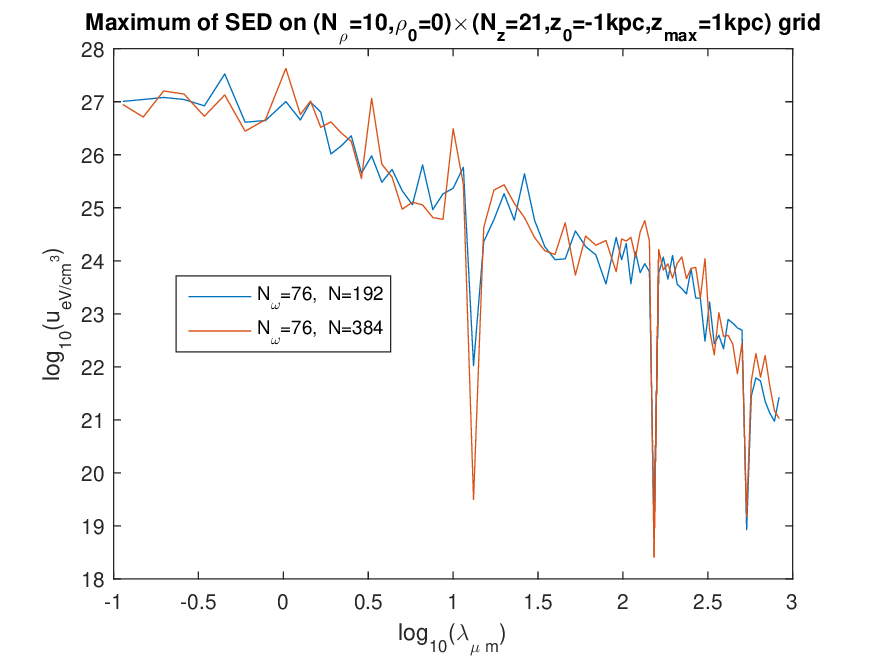}}
\caption{Effect of discretization number $N$ for: $N_\omega =76$, rough grid}
\label{76-rough}
\end{figure}

\begin{figure}[ht]
\centerline{\includegraphics[height=6cm]{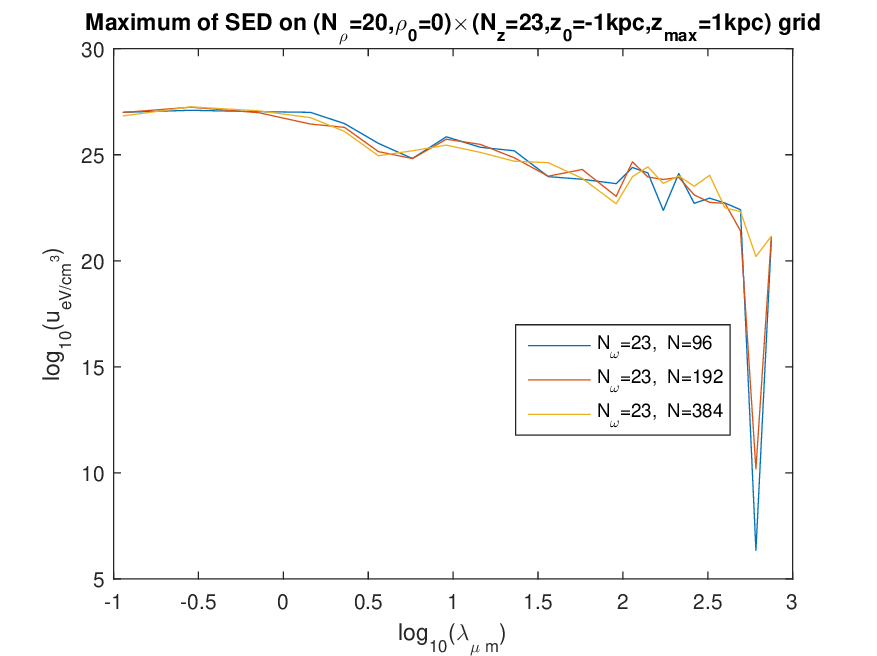}}
\centerline{\includegraphics[height=6cm]{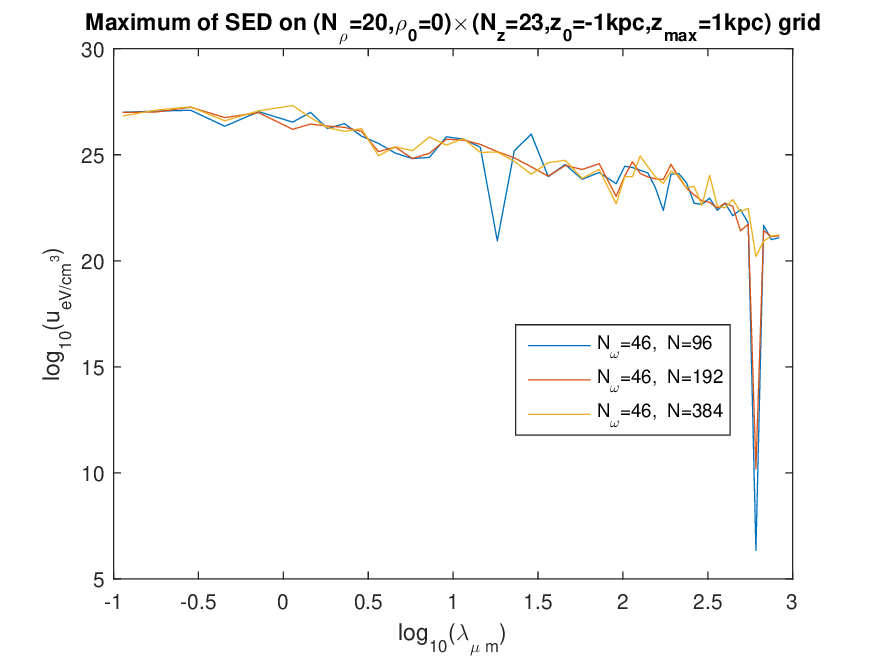}}
\centerline{\includegraphics[height=6cm]{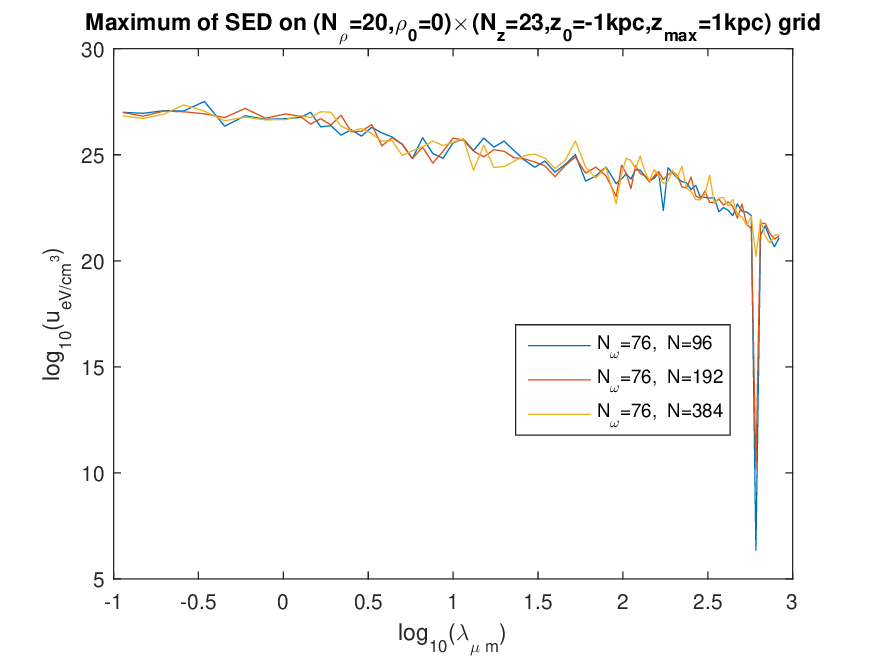}}
\caption{Effects of discretization number $N$ and number of frequencies $N_\omega$, fine grid}
\label{N&Nomega-fine}
\end{figure}

\begin{figure}[ht]
\centerline{\includegraphics[height=9cm]{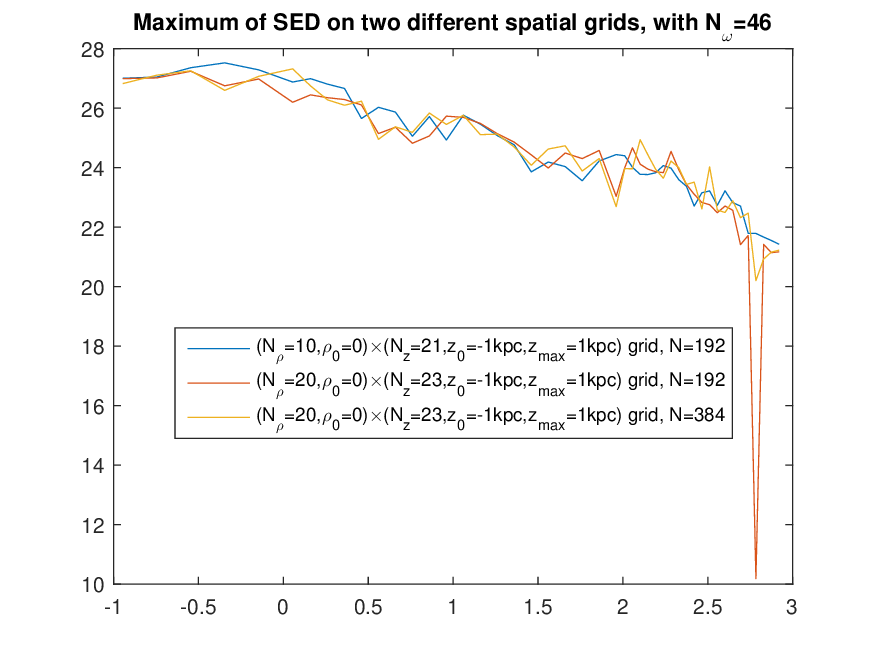}}
\caption{Comparison of two different grids}
\label{TwoGrids}
\end{figure}

\begin{figure}[ht]
\centerline{\includegraphics[height=9cm]{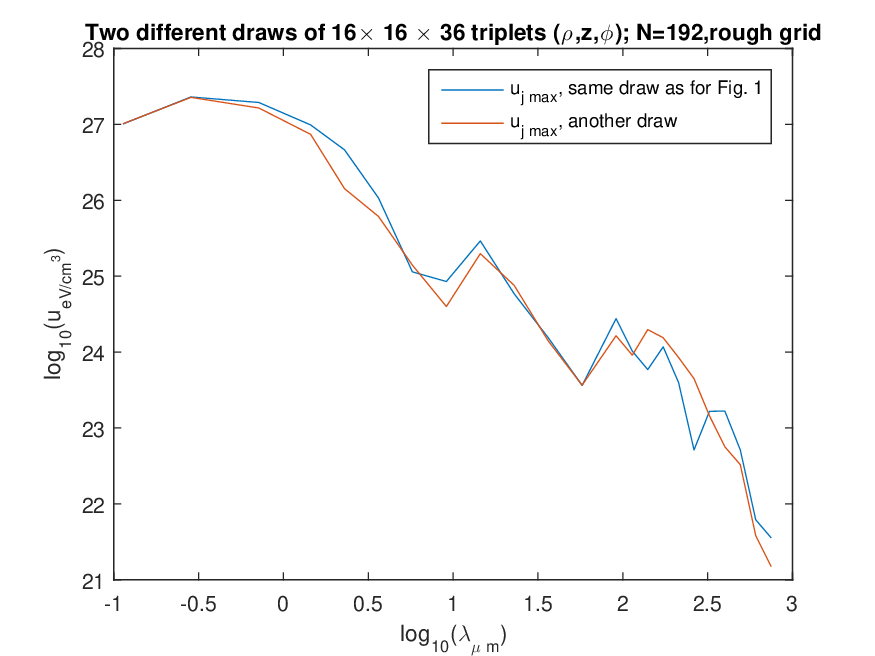}}
\caption{Comparison of two different draws of the set of ``stars"}
\label{TwoDraws}
\end{figure}

Figures \ref{23-rough} to \ref{N&Nomega-fine} show these effects. The most salient result is that {\it the extremely high values of $u_{j \mathrm{max}}$ are now found with all calculations and in the whole domain of $\lambda $} --- except that on some curves, abrupt oscillations toward lower values of the energy density are present for some wavelengths. By looking at the set of these figures, it is manifest that such abrupt oscillations occur when an inappropriate choice of parameters is done: essentially, the discretization number $N$ has to be large enough. (This is certainly expected, and this expectation is confirmed by the validation test in Ref. \cite{A61}.) Indeed, for a given value of $N_\omega $, those oscillations are maximum for the lowest value of $N$ in the trial ($N=12$) and progressively disappear when $N$ is increased. What is a ``large enough" value of $N$ is not strongly dependent of the fineness of the spatial grid (i.e., of whether the ``rough" one or the ``fine" one is used) and that of the frequency mesh ($N_\omega $). However, when using the finest frequency mesh ($N_\omega =76$) for the ``rough" spatial grid (Fig. \ref{76-rough}), increasing $N$ does not allow us to eliminate the abrupt oscillations toward lower values: it even happens then, that increasing $N$ from $192$ to $384$ actually deteriorates the $u_{j \mathrm{max}}=f(\lambda _j)$ curve. 
We interpret this as due to the fact that, when using a rougher spatial grid $G'$ for the fitting, less data are provided (the values taken  on the grid $G'$ by the l.h.s. of Eq. (\ref{Psi-simeq-Psi'-j-by-j-space})) to determine the unknowns $S_{nj}$ on the r.h.s. of (\ref{Psi-simeq-Psi'-j-by-j-space}) --- while, of course, increasing $N$ increases the number of unknowns and thus asks for more data. On the other hand, it is seen that (for the relevant values of $N$, say $N=192$ or $N=384$, so that little or no oscillations are present), the levels of $u_{j \mathrm{max}}$ depend quite little on $N_\omega $ i.e. on the fineness of the frequency mesh: compare the bottom figures between Figs. \ref{23-rough}, \ref{46-rough}, and \ref{76-rough}, and compare the three figures in Fig. \ref{N&Nomega-fine}. Also, the levels of $u_{j \mathrm{max}}$ depend quite little on whether the rough or the fine spatial grid is being used (see e.g. Fig. \ref{TwoGrids}). We also checked that the results are little dependent of the pseudo-random ``draw" of the set of point-like ``stars": another draw of $16\times 16 \times 36$ triplets $(\rho ,z,\phi )$ gives very similar curves  $u_{j \mathrm{max}} = f(\lambda _j)$ (Fig. \ref{TwoDraws}). In summary, we now find that, for the relevant values of $N$, say $N=192$ or $N=384$, $u_{j \mathrm{max}}$ decreases smoothly from $\simeq 10^{27}$ to $\simeq 10^{21} \mathrm{eV/cm}^3$ when $\lambda _j$ varies in the domain considered, i.e., from $\lambda \simeq 0.11\mu \mathrm{m} $ to $\simeq 830 \mu \mathrm{m}$. We note moreover that, for the low values of $\lambda_j $, the values of $u_{j \mathrm{max}}$ calculated using the present ``separate fitting" have the same (extremely high) magnitude as those calculated with the former ``grouped fitting" \cite{A62}. These observations lead us to conclude that: (i) the extremely high values of $u_{j \mathrm{max}}$ (in the whole domain of $\lambda $ considered) are really what the ``Maxwell model of the ISRF" predicts for this model of the Galaxy. (ii) Somewhat surprisingly, it is the {\it low} values of $u_{j \mathrm{max}}$ obtained for the higher values of $\lambda $ when the ``grouped fitting" was used \cite{A62} that were a numerical artefact. \\

\begin{figure}[ht]
\centerline{\includegraphics[height=9cm]{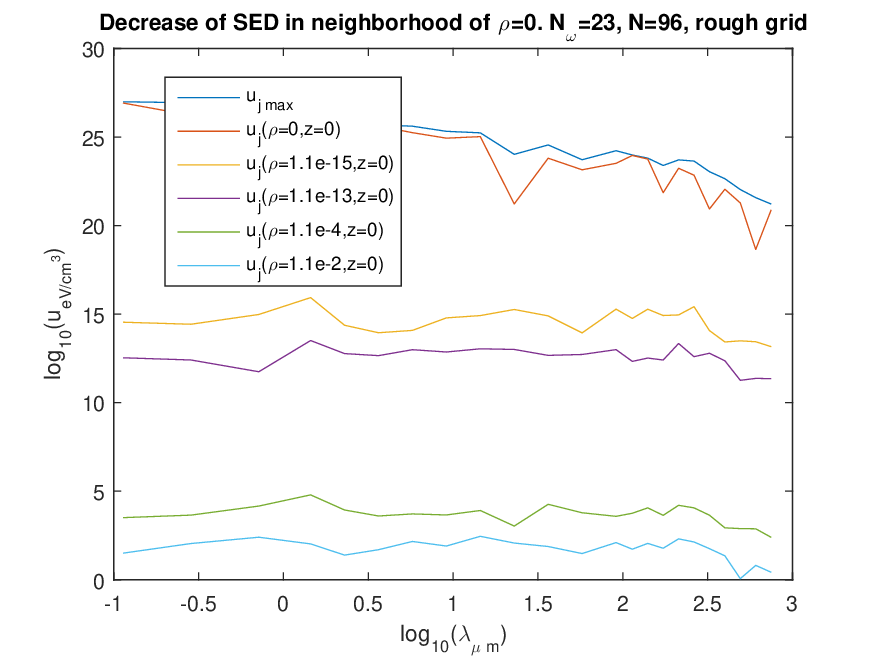}}
\caption{Decrease of the SED in the neighborhood of $\rho =0$. $\rho $ is given in kpc. For $-0.7 \lesssim \mathrm{log}_{10}\lambda \lesssim +0.7 $, the $u_{j \mathrm{max}}$ and $u_j(\rho=0,z=0)$ curves are hidden but nearly coincide. The $u_{j \mathrm{max}}$ curve is on Fig. 1 b).}
\label{Vois_rho_0}
\end{figure}

\subsection{Decrease of the energy density away from the axis}

Recall that the maxima of the $u_j$ 's, which are extremely high, are always obtained for $\rho =0$, i.e. on the axis of the (model of the) Galaxy, and that the energy density for $\rho =0$ depends little on $z$ in the domain investigated. The next questions are therefore: which is the extension around $\rho =0$ of the domain of the very high values? Do such values imply that ``too much EM energy" would be contained there? To answer these questions, we calculated the SED with successive lower and lower values of $\rho_\mathrm{max}$ (see Note \ref{def_grid}), starting from its value for the calculations shown on Figs. \ref{23-rough} to \ref{76-rough}, i.e., $\rho_\mathrm{max}=9$\,kpc, and decreasing it to $1, 10^{-1},...,10^{-14}$\,kpc, using the ``rough grid" parameters (see above), i.e., in particular, $N_\rho =10$, and using the $S_{n j}$ 's obtained with this rough grid with $\rho_\mathrm{max}=9$\,kpc --- so that, for $\rho_\mathrm{max}\ne 9$\,kpc, those calculations are not ones on the fitting grid. We looked in detail to the values $u_j(\rho _{m=2},z_{p=1}) = u_j(\rho _\mathrm{max}/9,z=0$). The main results are shown on Fig. \ref{Vois_rho_0}: even for very small values of $\rho \ne 0$, the values of $u_j$ are much smaller than $u_{j\,\mathrm{max}}$. That is, $u_j(\rho ,z)$ decreases very rapidly when $\rho $ is increased from 0. Actually, we found on the example of the smallest wavelength, corresponding with $j=1$, that, from $\rho =1$\,kpc \, down to \,$\rho =10^{-15}$\,kpc, we have to a good approximation \,$u_1(\rho,z=0 ) \simeq B/\rho $, with 
\be\label{B=}
B=u_1(\rho =1 \,\mathrm{kpc}, z=0)\simeq 10^{-0.45}\,(\mathrm{eV/cm}^3).\mathrm{kpc}.  
\ee
This behaviour is not valid until $\rho =0$, because 
for $\rho \rightarrow 0$, \ $u_1(\rho,z=0)$ tends towards $u_1(0,0)<\infty$, so we may assume $u_1(\rho,0) \lesssim B/\rho $. On the other hand, Fig. \ref{Vois_rho_0} shows that there is nothing special to $j=1$: we have $u_j \lesssim u_1$, moreover for $\rho \gtrsim 10^{-15}$\,kpc, $u_j$ depends only moderately on $\lambda _j$. We observed in our calculations that, for $\rho \le 1\,\mathrm{kpc}$, $u_j(\rho ,z)$ depends quite little on $z$ with $\abs{z} \le z_\mathrm{max}= 1\,$kpc. Thus we may give the following approximation (which is likely an overestimate) to $u_j$: for all $j$, and for\ $\abs{z} \le z_\mathrm{max}= 1\,$kpc, we have
\be\label{u_j=B/rho}
u_j(\rho ,z) \simeq B/\rho \quad \mathrm{for} \quad 10^{-15}\,\mathrm{kpc} \le \rho \le 1\,\mathrm{kpc} ,
\ee
\be\label{u_j<=B_j/rho} 
u_j(\rho,z ) \simeq u_j(\rho ) \lesssim B/\rho \ \ \mathrm{for}\ \ \rho  \le 10^{-15}\,\mathrm{kpc}, 
\ee 
with $u_j(\rho )$ a decreasing function of $\rho $. According to Eq. (\ref{u_j vs f}) of the Appendix, this implies that, for \ $\abs{z} \le z_\mathrm{max}= 1\,$kpc, we have also
\be\label{f=B_/rho}
f_{{\bf x}(\rho ,z)}(\lambda ) \simeq B/\rho \quad \mathrm{for} \quad 10^{-15}\,\mathrm{kpc} \le \rho \le 1\,\mathrm{kpc} ,
\ee
\be\label{f<=B_/rho}
f_{{\bf x}(\rho ,z)}(\lambda ) \lesssim B/\rho \quad \mathrm{for} \quad 0 \le \rho \le 10^{-15}\,\mathrm{kpc}, 
\ee
independently of $\lambda $ in the band 
\be\label{Lambda band}
\lambda ^{(1)}:=0.1\mu \mathrm{m} \le \lambda \le\lambda ^{(2)}:=830 \mu \mathrm{m}. 
\ee
With this approximation, we can assess the total EM energy (\ref{W_1_2_D-2}) contained in some disk 
\be\label{D(rho_1)}
D(\rho _1):\ (0 \le \rho \le \rho _1,\ 0\le \phi \le 2\pi,\ \abs{z} \le z_\mathrm{max}), 
\ee
with $\rho _1\leq 1\,\mathrm{kpc}$, and in the wavelength band $[\lambda ^{(1)},\lambda ^{(2)}]$. This energy is bounded, owing to (\ref{f=B_/rho})--(\ref{f<=B_/rho}), by
\be
W_{1-2,\,D(\rho_1)} \lesssim \mathrm{Log} \frac{\lambda ^{(2)}}{\lambda ^{(1)}}\times \int _{D(\rho_1)} \ \frac{B}{\rho({\bf x})} \dd ^3 {\bf x} = \mathrm{Log}\frac{\lambda ^{(2)}}{\lambda ^{(1)}}\times  \int _{D(\rho_1)} \ \frac{B}{\rho} \rho \,\dd \rho  \,\dd \phi \,\dd z  ,
\ee
i.e.
\be\label{W_small_rho_le}
W_{1-2,\,D(\rho_1)} \lesssim \mathrm{Log} \frac{\lambda ^{(2)}}{\lambda ^{(1)}}\times  B \,\rho _1 \times 2\pi \times 2z_\mathrm{max}.
\ee
But consider, instead of the disk $D(\rho _1)$, the ring  $R(\rho _0,\rho _1): \ (\rho _0 \le \rho \le \rho _1,\ 0\le \phi \le 2\pi,\ \abs{z} \le z_\mathrm{max})$, with $\rho _0 \ge 10^{-15}\,\mathrm{kpc}$. (Thus a ring with a very narrow aperture.) Using this time only (\ref{f=B_/rho}), the same calculation gives 
\be\label{W_small_rho_hole}
W_{1-2,\,R(\rho _0,\rho_1)} \simeq \mathrm{Log} \frac{\lambda ^{(2)}}{\lambda ^{(1)}}\times  B \,(\rho _1-\rho _0) \times 2\pi \times 2z_\mathrm{max}.
\ee
Taking $\rho _0 = 10^{-15}\,\mathrm{kpc}$, the conjunction of (\ref{W_small_rho_le}) and (\ref{W_small_rho_hole}) shows that the contribution of the domain with $0 \le \rho \le \rho _0$ is totally negligible, hence we may write
\be\label{W_small_rho}
W_{1-2,\,D(\rho_1)} \simeq \mathrm{Log} \frac{\lambda ^{(2)}}{\lambda ^{(1)}}\times  B \,\rho _1 \times 2\pi \times 2z_\mathrm{max}.
\ee
We can calculate the contribution $\delta U$ that it gives to the average density of the EM energy in some disk  \,$D(\rho _2)$ of the Galaxy, with $\rho _2\ge \rho _1$, making a volume $V_2=\pi \rho _2^2 z_\mathrm{max}$:
\be\label{delta U}
\delta U :=\frac{W_{1-2,\,D(\rho_1)}}{V_2} \simeq 4\,\mathrm{Log} \frac{\lambda ^{(2)}}{\lambda ^{(1)}}\times \frac{  B \,\rho_1}{ \rho _2^2}.
\ee
(Note that we may leave $B$ in $(\mathrm{eV/cm}^3).\mathrm{kpc}$ and $\rho _1$ and $\rho _2$ in kpc.) To give figures, let us first take $\rho _1=\rho _2=1\, \mathrm{kpc}$, so that the corresponding value of $\delta U$ is just the average volumic energy density $\langle U \rangle _{D(1\,\mathrm{kpc})}$ in the disk $D(\rho _1=1\,\mathrm{kpc})$. Then (\ref{delta U}) with (\ref{B=}) give us 
\be\label{Ubar_1kpc}
\langle U \rangle _{D(1\,\mathrm{kpc})} \simeq 51 \,\mathrm{eV/cm}^3. 
\ee
Note that this value is {\it not} very high. Another interesting application of Eq. (\ref{delta U}) is to assess the effect, on that average value in the same domain $D(1\,\mathrm{kpc})$, of the domain of the ``very high" values of the SED, say the domain for which $u_j \geq 10^{6}\,\mathrm{eV/cm}^3$ --- i.e., from (\ref{B=}) and (\ref{u_j=B/rho}), $\rho \le \rho _\mathrm{vh}$, with 
\be\label{rho_vh}
\rho _\mathrm{vh} = 10^{-6.45} \simeq 3.55\times 10^{-7} \,\mathrm{kpc} \simeq 1.1 \times 10^{10} \mathrm{\,km}, 
\ee 
which is almost twice the average distance Sun-Pluto, but still very small on a galactic scale. Taking to this effect $\rho _1=\rho _\mathrm{vh}$ and $\rho _2=1\,\mathrm{kpc}$ in Eq. (\ref{delta U}), the numerical values (\ref{B=}), (\ref{Lambda band}) and (\ref{rho_vh}) give us $\delta U \simeq 4.54 \times 10^{-6} \,\mathrm{eV/cm}^3$. {\it In summary,} the ``very high" values of the SED are confined to the close neighborhood of the Galaxy's axis and contribute negligibly to the average energy density (\ref{Ubar_1kpc}) in the disk $D(1\,\mathrm{kpc})$.

\section{Results: spatial variation of the SED \& comparison with the literature}\label{Field_of_SED}

This model's prediction for the spatial variation of the SED in the Galaxy was investigated, using again the separate fitting and the adjustment of the local SED on the measured values (both being described in Sect. \ref{SeparateFitting}). It was shown by using two different types of representations. \\

\begin{figure}[ht]
\centerline{\includegraphics[height=12cm]{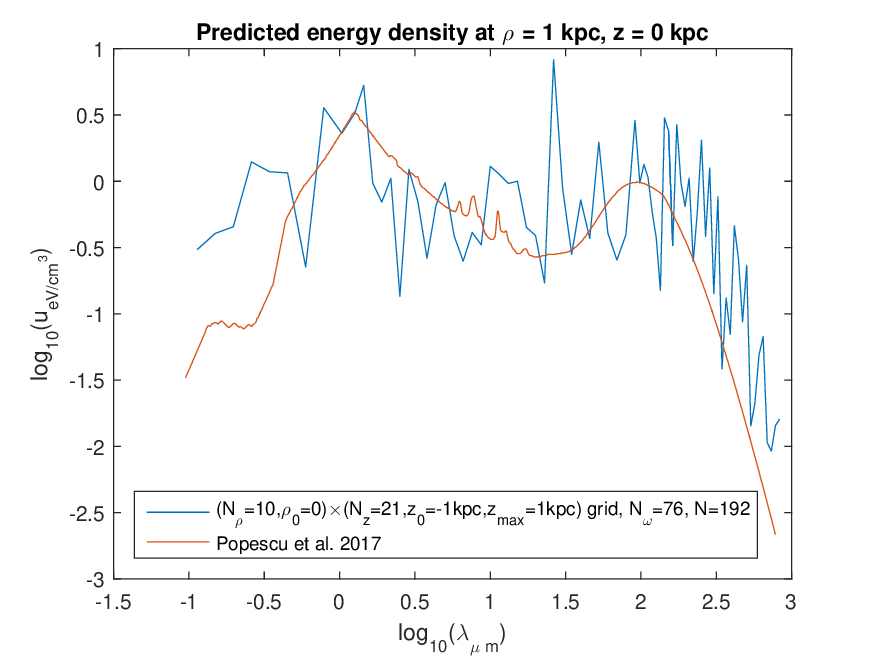}}
\caption{SED at ($\rho = 1$\,kpc, $z = 0$)}
\label{SED(1 0)}
\end{figure}

\begin{figure}[ht]
\centerline{\includegraphics[height=12cm]{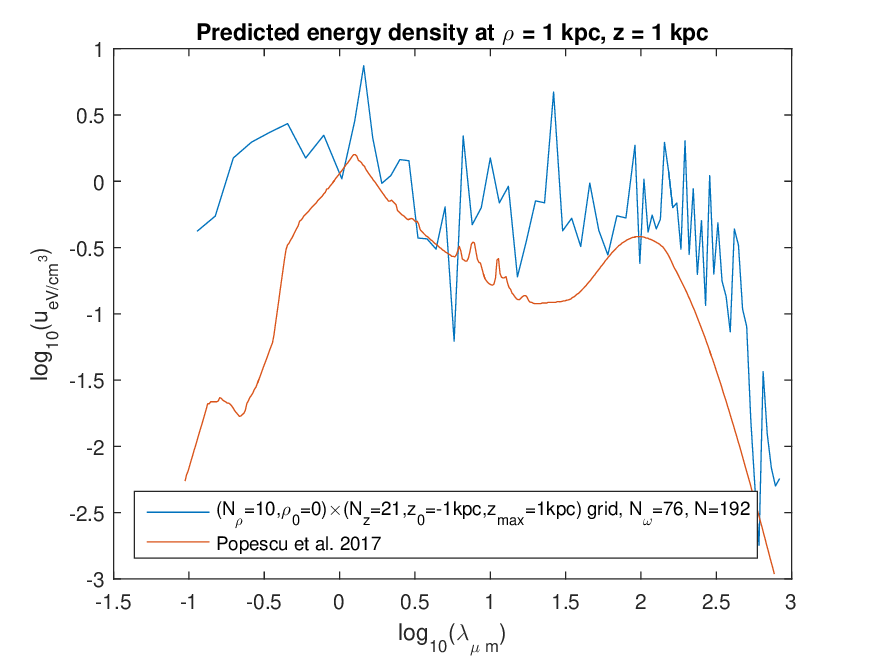}}
\caption{SED at ($\rho = 1$\,kpc, $z = 1$\,kpc)}
\label{SED(1 1)}
\end{figure}

\begin{figure}[ht]
\centerline{\includegraphics[height=12cm]{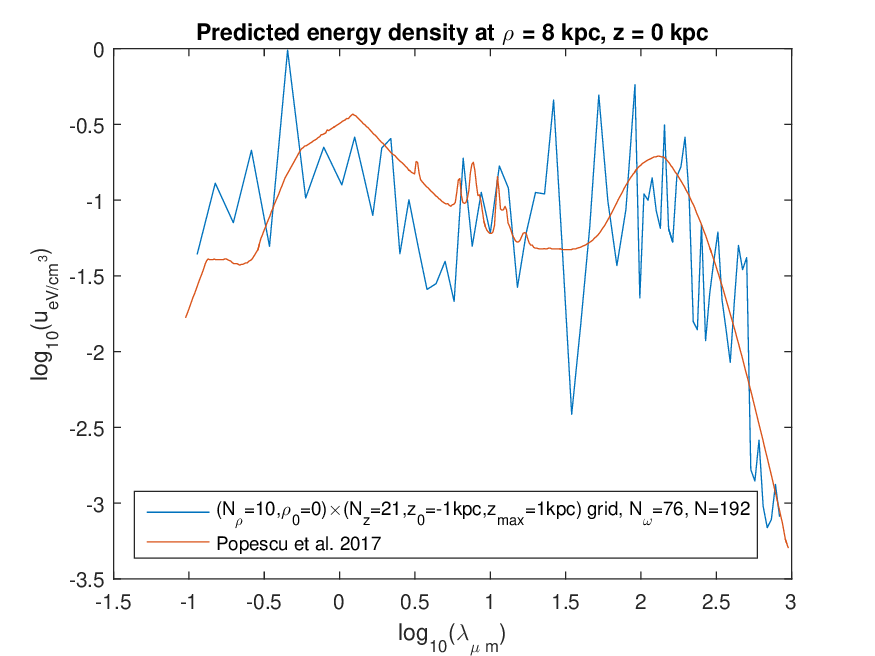}}
\caption{SED at ($\rho = 8$\,kpc, $z = 0$)}
\label{SED(8 0)}
\end{figure}

\begin{figure}[ht]
\centerline{\includegraphics[height=12cm]{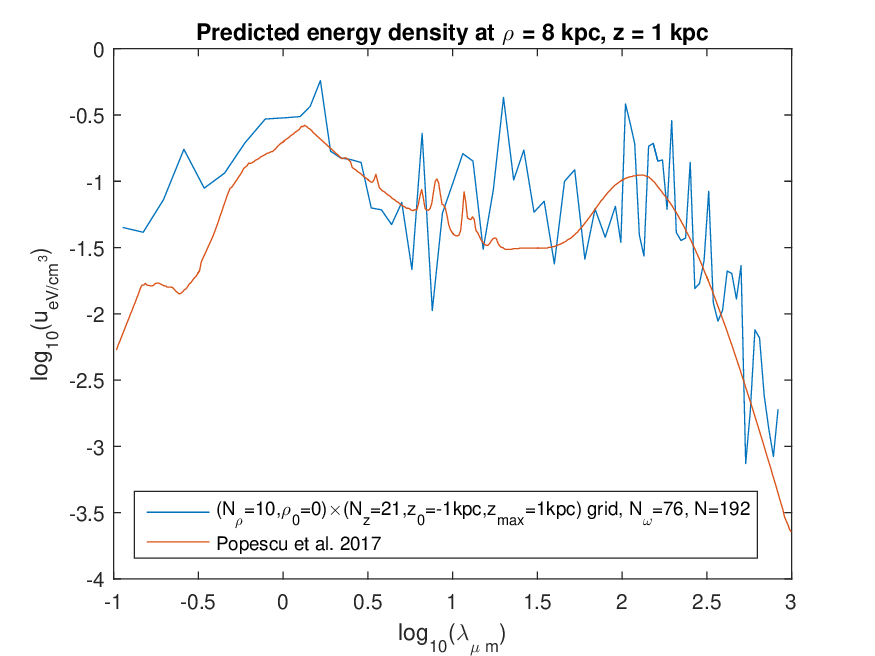}}
\caption{SED at ($\rho = 8$\,kpc, $z = 1$\,kpc)}
\label{SED(8 1)}
\end{figure}

First, we plotted the SED at four different points in the Galaxy, and we compared the results with those obtained by Popescu {\it et al.} \cite{Popescu-et-al2017}, who used a radiation
transfer model built by them. (Their model also assumes axisymmetry.) Figures \ref{SED(1 0)}--\ref{SED(8 1)} show this comparison, our model being used here with $N=192$ and $N_\omega =76$. (Other choices of parameters that we tried gave similar figures.) It can be seen that the predictions of the present model do not differ very significantly from those of the radiation transfer model of Ref. \cite{Popescu-et-al2017}. The main difference is that our calculations oscillate somewhat strongly with the wavelength. The comparison of Figs. \ref{SED(1 0)}--\ref{SED(8 1)} here with Figs. 2-5 in Ref. \cite{A62} shows that the difference between the results of the two models is significantly smaller now than it was in our previous work, in which the calculations were based on the grouped fitting \cite{A62}: the difference in $\mathrm{log}_{10} (u_j)$ between the results of our model and Ref. \cite{Popescu-et-al2017} is here $\lesssim 1$, whereas it went beyond $3$ and even $4$ in the previous calculations. However, the new calculations oscillate with the wavelength also at higher wavelengths. Whereas, when the grouped fitting was used, there was virtually no oscillation for $\lambda \gtrsim 10 \,\mu \mathrm{m}$ at the two positions at $\rho =8\,\mathrm{kpc}$. (There were oscillations in the whole range of $\lambda $ for the two positions at $\rho =1\,\mathrm{kpc}$.) In order to check if those calculations inside the spatial domain of small values of the SED could be ``polluted" by the extremely high values on the galaxy's axis, we investigated the effect of doing the fitting on a ``shifted" grid with $\rho \ge 1\,\mathrm{kpc}$. This did not lead to less oscillations. The general reason for these oscillations may be simply that this model takes fully into account the wave nature of the EM field.\\

\begin{figure}[ht]
\centerline{\includegraphics[height=8cm]{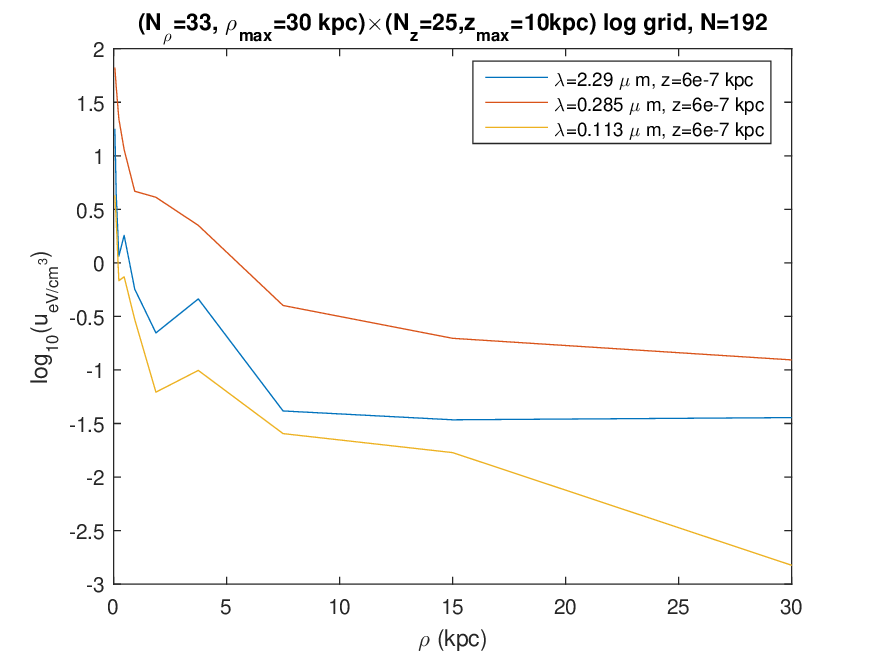}}
\centerline{\includegraphics[height=8cm]{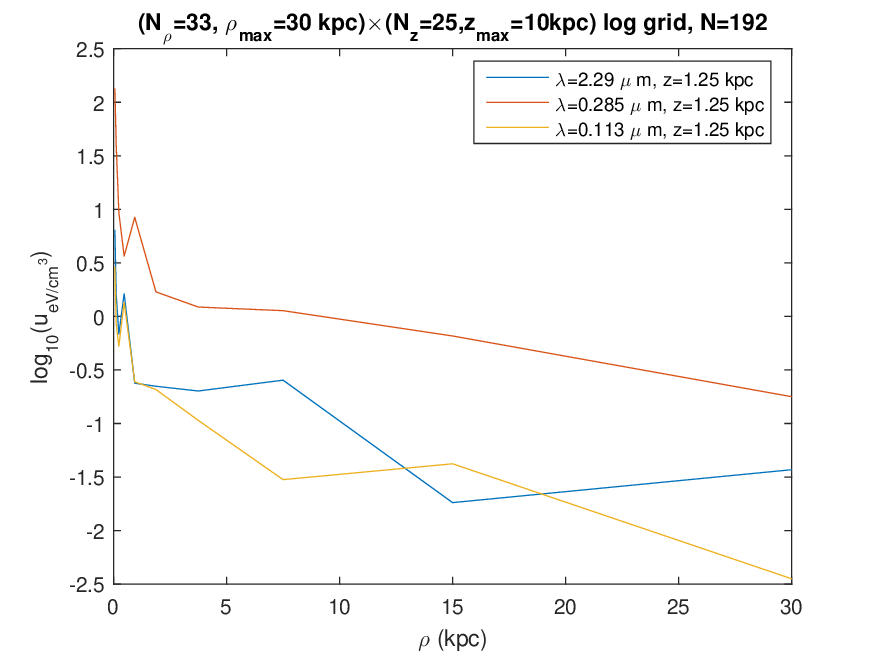}}
\caption{Radial profiles of radiation fields. Fitting done on a logarithmic grid: $\rho _1 = \rho _\mathrm{max}, \rho _m=\rho _{m-1}\times q\ (m=2,...,N_\rho)$; $z _1 = z _\mathrm{max}, z _k=z _{k-1}\times q\ (k=2,...,N_z)$; $q=0.5$. SED values at $(\rho _m,z_{N_z})\ (m=1,...,10 )$, then at  $(\rho _m,z_4)\ (m=1,...,10 )$, are plotted.}
\label{K-B-UV_z_0_1_log}
\end{figure}

\begin{figure}[ht]
\centerline{\includegraphics[height=8cm]{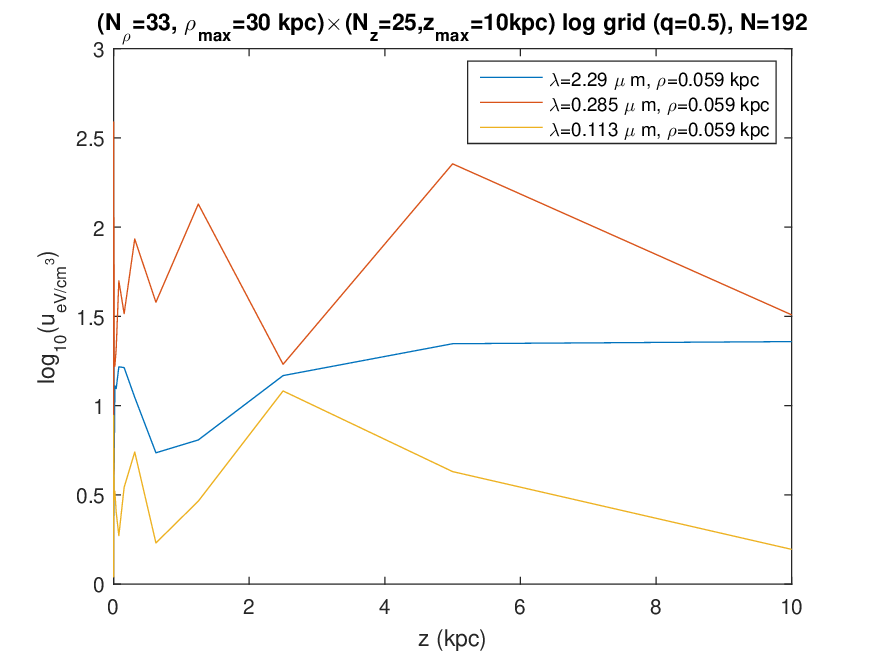}} 
\centerline{\includegraphics[height=8cm]{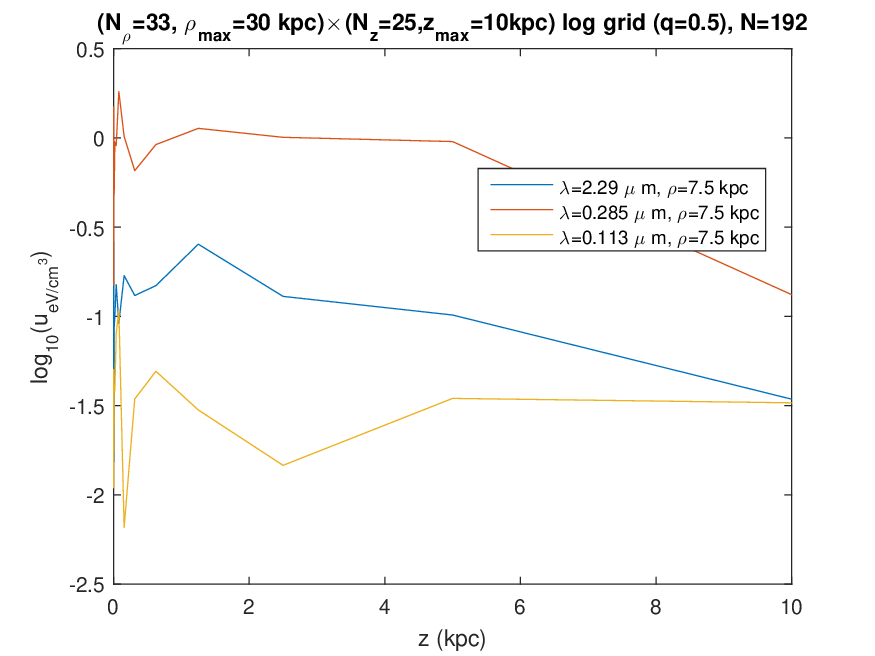}}
\caption{Vertical profiles of radiation fields. Fitting done on the same logarithmic grid as for Fig. \ref{K-B-UV_z_0_1_log}. SED values at $(\rho _{10},z_k)\ (k=1,...,N_z)$, then at $(\rho _3,z_k)\ (k=1,...,N_z)$, are plotted.}
\label{K-B-UV_rho_0_8_log}
\end{figure}

\begin{figure}[ht]
 \centerline{\includegraphics[height=8cm]{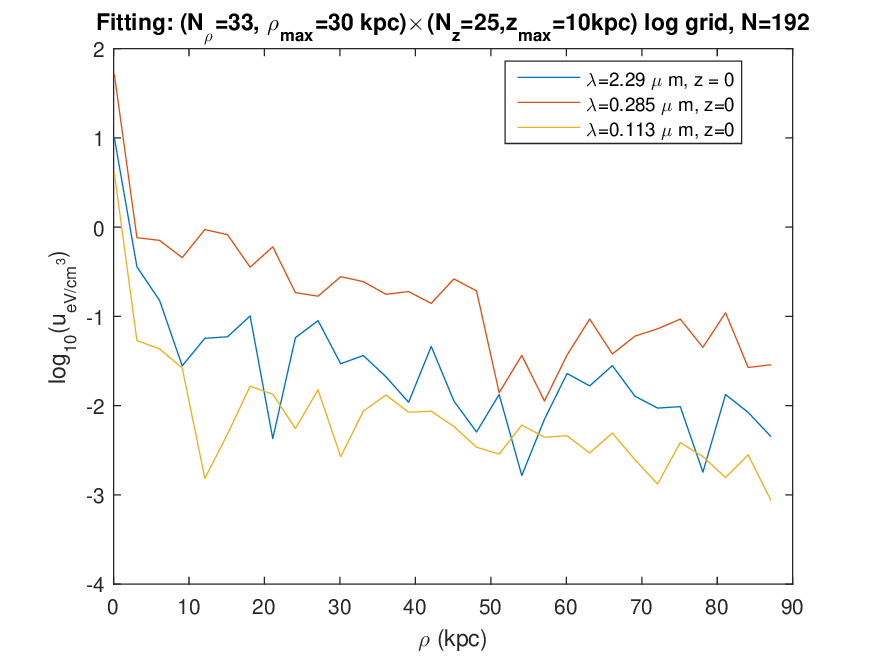}}
 \centerline{\includegraphics[height=8cm]{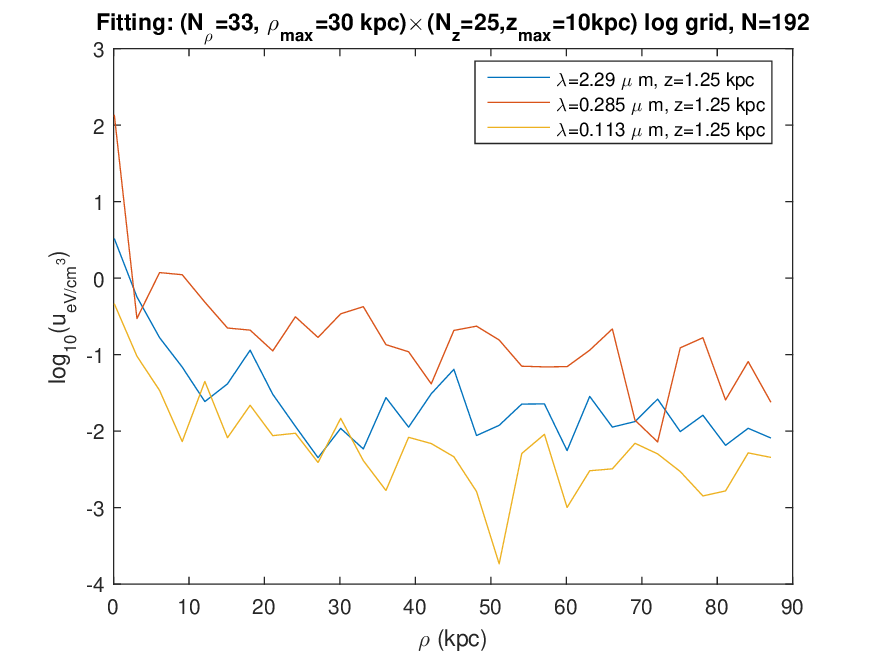}}

\caption{Radial profiles of radiation fields. Fitting done on the same logarithmic grid as for Fig. \ref{K-B-UV_z_0_1_log}. SED values at regularly spaced values of $\rho$, starting at $0.1\,\mathrm{kpc}$, and for $z=0$, then $z=1.25\,\mathrm{kpc}$, are plotted.}
\label{K-B-UV_z_0_1_reg}
\end{figure}

\begin{figure}[ht]
\centerline{\includegraphics[height=8cm]{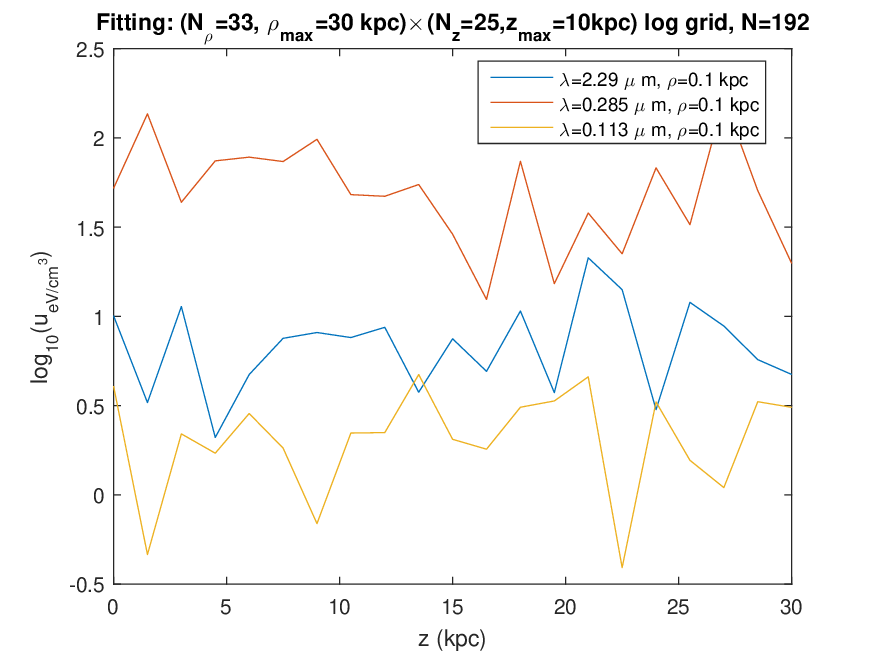}}
\centerline{\includegraphics[height=8cm]{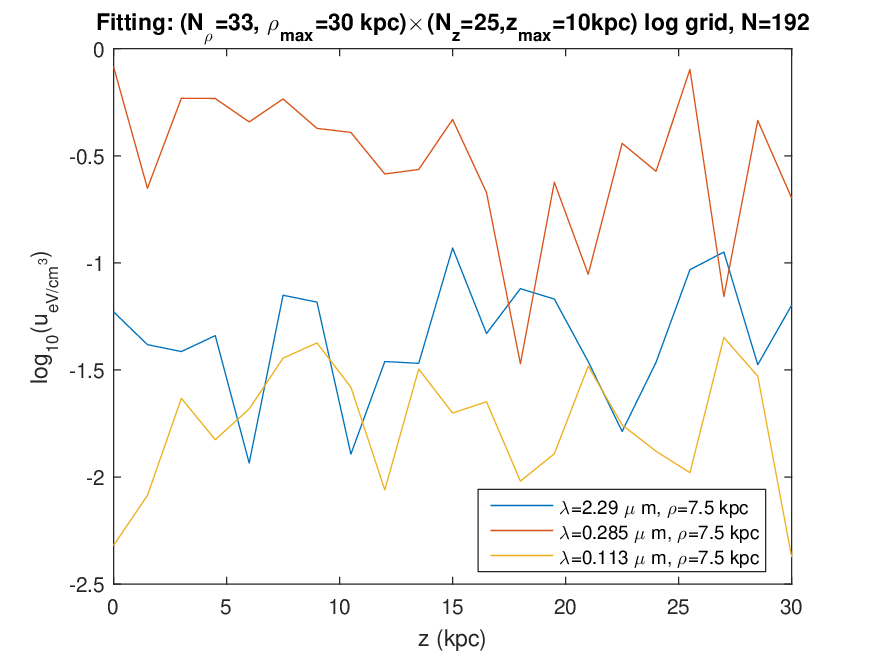}}
\caption{Vertical profiles of radiation fields. Fitting done on the same logarithmic grid as for Fig. \ref{K-B-UV_z_0_1_log}. SED values at regularly spaced values of $z$, starting at $0$, and for $\rho =0.1$, then $\rho =7.5\,\mathrm{kpc}$, are plotted.}
\label{K-B-UV_rho_0_8_reg}
\end{figure}

Second, we plotted the radial and vertical profiles of the radiation fields at three wavelengths close to the ones considered in Fig. 7 of Popescu {\it et al.} \cite{Popescu-et-al2017} (``K, B , UV"). Figures \ref{K-B-UV_z_0_1_log} and \ref{K-B-UV_rho_0_8_log} show these profiles as they are calculated at points $(\rho ,z)$ belonging to the ``logarithmic" grid on which the fitting was done for this calculation (see the legend). Those profiles of the radiation fields obtained with the present model on the fitting grid are relatively similar to those that are predicted with the very different model of Ref. \cite{Popescu-et-al2017}, both in the levels and in the rough shape of the profiles. The most important difference is seen on the vertical profiles of Fig. \ref{K-B-UV_rho_0_8_log}: according to the Maxwell model of the ISRF, the energy density level decreases more slowly when the altitude $z$ increases --- or even, for the $\lambda =2.29 \mu \mathrm{m}$ radiation at $\rho =0.059\,$kpc or the $\lambda =0.113 \mu \mathrm{m}$ radiation at $\rho =7.5\,$kpc, the level of the SED does not decrease in the range considered for $z$. A similar lack of decrease is found on the radial profiles of Fig. \ref{K-B-UV_z_0_1_log}, for the $\lambda =2.29 \mu \mathrm{m}$ radiation, either at $z\simeq 0$ or at $z=1.25 \,\mathrm{kpc}$. Using that same fitting done on a logarithmic grid, we also calculated and plotted the radial and vertical profiles of the same radiations, but this time for regularly spaced values of $\rho $ (or respectively $z$), and in a larger range for $\rho $ (or respectively $z$), Figs. \ref{K-B-UV_z_0_1_reg} and \ref{K-B-UV_rho_0_8_reg}. The radial profiles of Figs. \ref{K-B-UV_z_0_1_log} and \ref{K-B-UV_z_0_1_reg} are consistent, although, in contrast with Fig. \ref{K-B-UV_z_0_1_log}, Fig. \ref{K-B-UV_z_0_1_reg} plots the SED at points $(\rho ,z)$ which were not involved in the fitting, and which moreover involve an extrapolation to a larger range for $\rho $ as compared with the fitting. The vertical profiles of Fig. \ref{K-B-UV_rho_0_8_reg}, which also correspond with points which were not involved in the fitting, and also involve an extrapolation to a larger range for $z$ as compared with the fitting, show an oscillating behaviour without any tendency to a decrease at large $z$. 


\section{Asymptotic behaviour at large $\rho $ and at large $z$}\label{Asymptotic}

To help understanding the behaviours just noted, in this section we study the asymptotic behaviour of the expressions of the components of the EM field and of the SED, as they are given by the Maxwell model of the radiation field. The expressions (\ref{Bphi'})--(\ref{Ez'}) that are implemented in the numerical model, are deduced from the exact integral expressions of the EM field for a given angular frequency $\omega $, after the summation over the frequencies, and after the discretization (\ref{k_nj}) is done. Hence, we begin with the exact integral expressions of the EM field for a given angular frequency. These expressions, which are valid for any totally propagating, axisymmetric, time-harmonic EM field, are (Eqs. (13)--(15) in Ref. \cite{A61}):
\be\label{Bphi_mono}
B_{\phi \,\omega \, S} = {\mathcal Re} \left[e^{-\iC  \omega t} \int_{-K} ^{+K} \sqrt{K^2-k^2}\, J_1\left(\rho \sqrt{K^2-k^2}\right ) \,S(k) \,e^{\iC kz} \dd k \right ],
\ee
\be\label{Erho_mono}
E_{\rho \, \omega \, S} = {\mathcal Re} \left[-\iC \frac{c^2}{\omega } e^{-\iC  \omega t} \int_{-K} ^{+K} \sqrt{K^2-k^2}\, J_1\left(\rho \sqrt{K^2-k^2}\right )\iC k \,S(k) \,e^{\iC kz} \dd k \right ],
\ee
\be\label{Ez_mono}
E_{z \, \omega \, S} = {\mathcal Re} \left[\iC e^{-\iC \omega t} \int_{-K} ^{+K} J_0\left(\rho \sqrt{K^2-k^2}\right )\,\left(\omega -\frac{c^2}{\omega }\,k^2 \right )\,S(k)\,e^{\iC kz} \dd k \right ],
\ee
where $K := \omega /c$ --- the other components being obtained by the duality (\ref{dual}) from the components (\ref{Bphi_mono})--(\ref{Ez_mono}), with in the most general case an other spectrum function $S'(k)$. \\

The dependence in $\rho $ of the components (\ref{Bphi_mono})--(\ref{Ez_mono}) is determined by that of the Bessel functions $J_0$ and $J_1$, and by the form of the integrals which involve them. At large $x$ we have the asymptotic expansion \cite{Dieudonne_CI}
\be\label{expans_J_large_x}
J_\alpha(x )= \sqrt{\frac2{\pi x }}\cos \left( x -\frac{\alpha\pi}2- \frac{\pi}4\right) + O\left(x ^{-\frac{3}{2}} \right ).
\ee
However, the argument of the Bessel functions in Eqs. (\ref{Bphi_mono})--(\ref{Ez_mono}) is $x= \rho \sqrt{K^2-k^2}$. Hence, as $\rho \rightarrow \infty $, $x$ does not tend towards $\infty $ uniformly, depending on the integration variable $k$: we even have $x\equiv 0$ independently of $\rho $, for $k=\pm K$. Therefore, it is not obvious to see if the integrals (\ref{Bphi_mono})--(\ref{Ez_mono}) do have an expansion at fixed $z$ as $\rho \rightarrow \infty$. \\

As to the behaviour at fixed $\rho $ and at large $z$: up to the real part, and for a fixed value of $\rho $, the components (\ref{Bphi_mono})--(\ref{Ez_mono}) are expressions of the form $e^{-\iC \omega t} I(z)$, with
\be\label{Integral_phase}
I(z) = \int_a ^b f(k) e^{\iC z g(k)}\, \dd k,
\ee
and where, specifically,  $\,a=-K,\, b=+K$, and the phase function is simply $\ g(k)\equiv k$, which has no stationary point. (The regular function $k\mapsto f(k)$ depends on the component being considered, and also on $\rho $ as a parameter.) In that case, we have the asymptotic expansion \cite{WikiPhaseStationn}
\be\label{Expans_Integral_phase_2}
I(z) = \frac{f(K)}{\iC z }\mathrm e^{\iC z K} -  \frac{f(-K)}{\iC z }\mathrm e^{-\iC z K} + O\left(\frac1{z^2}\right).
\ee
So at large $z$ and for a fixed value of $\rho $, all components of any totally propagating, axisymmetric, time-harmonic EM field are order $\frac{1}{z}$ (unless the coefficient of $\frac{1}{z}$ in this expansion is zero, which is true only in particular cases --- then the relevant component is higher order in $\frac{1}{z}$). This applies indeed to the part ({\bf i}) of the decomposition \hyperlink{GAZR1}{({\bf i})-({\bf ii})}, that is given by Eqs. (\ref{Bphi_mono})--(\ref{Ez_mono}), but also to the part ({\bf ii}), since it is obtained from (\ref{Bphi_mono})--(\ref{Ez_mono}) by applying the EM duality (\ref{dual}) (with, in the most general case, a different spectrum function $S'(k)$). Hence {\it the SED (\ref{Udiscrete}) is order $\frac{1}{z^2}$ at large $z$,} for any fixed value of $\rho $ --- when the $C^{(q)}_j({\bf x})$ coefficients correspond with the exact expressions (\ref{Bphi_mono})--(\ref{Ez_mono}). [The explicit expression of the coefficient of $\frac{1}{z^2}$, depending on $\rho $ , $K$, $S(K)$, $S(-K)$ (and, in the most general case, of the values $S'(K)$, $S'(-K)$ of the spectrum function $S'$ corresponding to the part ({\bf ii}) of the decomposition \hyperlink{GAZR1}{({\bf i})-({\bf ii})}) might easily be obtained from (\ref{Udiscrete}), (\ref{Bphi_mono})--(\ref{Ez_mono}), and (\ref{Expans_Integral_phase_2}).] The foregoing result applies to a general spectrum function $S(k)$ (and $S'(k)$). By summation on the frequency index $j$, it extends to an EM field having a finite set of frequencies. \\

Let us now investigate the asymptotic behaviour of the EM field and the SED
, still in the totally propagating case with axial symmetry, but now after the summation over the frequencies and the discretization (\ref{k_nj}). After the discretization, each among the  $C^{(q)}_j$ coefficients in the expansions (\ref{F(t)}) of the components of the EM field has the form \cite{A62}:
\be\label{C^q_j}
C^{(q)}_j = C^{(q)}_j(\rho ,z) = \sum _{n=0} ^N R'\,_n^{(q)} \,J_\alpha \left(\rho \frac{\omega _j}{\omega _0} R_n \right) G_{nj}(z)\qquad (\alpha =0\ \mathrm{or}\ \alpha =1),
\ee
where $R'\,_n^{(q)} > 0\ $ (except for $R'\,_0^{(q)}$ and $R'\,_N^{(q)}$, both of which turn out to be zero) are constant numbers, and where $G_{n j}(z) = \exp \left( \iC \omega _j\, t \right )  F_{n j}(t,z)$ is just the function $F_{n j}$ in Eq. (\ref{F_nj}) hereabove, deprived of its periodic time-dependence (and thus is a periodic function of $z$). Together with (\ref{expans_J_large_x}), Eq. (\ref{C^q_j}) shows that, at a given value of $z$, we have $C^{(q)}_j=O(1/\sqrt{\rho })$ as $\rho \rightarrow \infty $. The SED for an EM field having a finite set of frequencies is given by Eq. (\ref{Udiscrete}). For any given frequency $(j)$, $u_j$ is a quadratic form of the $C^{(q)}_j$ coefficients, hence
\be
u_j(\rho ,z) = O\left(\frac{1}{\rho } \right )\quad (\rho \rightarrow \infty ).
\ee
This is compatible with the curves shown on Fig. \ref{K-B-UV_z_0_1_reg}.

Passing to the behaviour at large $z$: in Eq. (\ref{C^q_j}), the dependence in $z$ is entirely contained in the functions $G_{n j}(z)$ which, we noted, are periodic. Hence, the coefficients $C^{(q)}_j(\rho ,z)$, each of which involves a linear combination of these functions (with coefficients that depend on $\rho$), are {\it almost-periodic functions of $z$} \cite{Ameriosi-Prouse1971}, and the same is true for the components (\ref{F(t)}) of the EM field. Moreover, for any given value of $\rho $, each $u_j$ in Eq. (\ref{Udiscrete}) is hence the sum of the square moduli of periodic complex functions of $z$. Therefore \cite{Ameriosi-Prouse1971}, {\it the SED is an almost-periodic function of $z$, too.} This result allows us to understand the lack of a decrease with $z$, observed on the vertical profiles of Fig. \ref{K-B-UV_rho_0_8_reg}, which involve an extrapolation to a larger range for $z$ as compared with the domain used for the fitting: an almost-periodic function $f$ does not tend towards zero at infinity, unless $f\equiv 0$.
\footnote{\ 
This results from the most common definition of an almost-periodic function $f$ \cite{Ameriosi-Prouse1971}: the existence, for any $\epsilon >0$, of a relatively dense set of $\epsilon $ almost-periods. I.e., for any $\epsilon >0$,  there exists a length $l_\epsilon $ such that, for any $x \in \mathbb{R}$, there is at least one number $T \in [x,x+l_\epsilon [$ such that for any $t\in \mathbb{R}$, \ $\abs{f(t+T)-f(t)} \le \epsilon $. If $f$ is not identically zero, let $a \in \mathbb{R}$, such that $f(a)=\alpha \ne 0$. Taking $\epsilon =\frac{\alpha }{2}$ in the definition above, we thus have $\abs{f(a+T)-f(a)} \le \frac{\alpha }{2} $, hence $\abs{f(a+T)} \ge \frac{\alpha }{2}$. Since $x$ can be taken arbitrarily large and since $T\ge x$, this proves that $f$ does not tend towards zero at infinity.
}
As to Figs. \ref{K-B-UV_z_0_1_log} and \ref{K-B-UV_rho_0_8_log}, they involve no extrapolation, thus the relevant coefficients result from the fitting done on the very domain to which the curves belong. Hence the asymptotic behaviour of $u_j$ (whether at large $z $ or at large $\rho $) is not relevant to them.

\section{Discussion and conclusion}\label{Conclusion}

In this paper, we developed an improved numerical scheme to adjust the Maxwell model of the ISRF in a galaxy, which was proposed in a foregoing work \cite{A61}. Namely, at the stage of fitting the radiations emitted by the many different point-like ``stars" which make the model galaxy, we are now considering each time-harmonic component separately, which is more precise. This allows us as a bonus to eliminate the time variable at this stage, Eq. (\ref{Psi-simeq-Psi'-j-by-j-space}) --- thus reducing the computer time. \\

We used that ``separate fitting" procedure, first, to check if the extremely high values of the spectral energy density (SED), which were predicted by this model on the axis of our Galaxy with the former ``grouped fitting" \cite{A62}, are a physical prediction or a numerical artefact. A rather detailed investigation led us to conclude that these extremely high values are indeed what the model predicts --- see Sect. \ref{Robust high}. However, we find also that the SED decreases very rapidly when one departs from the galaxy's axis, see Fig. \ref{Vois_rho_0}. Moreover, the average energy density of the EM field in, for example, a disk of diameter $1\,\mathrm{kpc}$ and thickness $2\,\mathrm{kpc}$, is not very high, Eq. (\ref{Ubar_1kpc}). The extremely high values of the SED on the axis of our Galaxy (and likely also in many other galaxies) are a new and surprising prediction for the ISRF. Recall that our model is adjusted so that the SED predicted for our local position in the Galaxy coincide with the SED determined from spatial missions, and thus is fully compatible with what we see of the ISRF from where we are. The prediction of the present model may be interpreted as a kind of self-focusing of the EM field in an axisymmetric galaxy. On the other hand, as we mentioned in the Introduction, the existing (radiation-transfer) models for the ISRF do not consider the EM {\it field} with its six components coupled through the Maxwell equations. These models consider paths of photons or rays and do not take into account the nature of the EM radiation as a field over space and time, subjected to specific PDE's. So a self-focusing cannot be seen with those models. It is difficult to assess the degree to which this prediction depends on the specific assumptions of the model, in particular the axial symmetry. If this prediction is at least partly confirmed, this will have important implications in the study of galaxies.\\

Second, we studied the spatial variation of the SED predicted by our model with the new procedure, and compared it with the predictions of a recent radiation transfer model \cite{Popescu-et-al2017}. The difference between the results of the two models is much smaller now than it was \cite{A62} with the older procedure. However, the SED predicted by our model still oscillates as function of the wavelength (or the frequency) also with the new, ``separate fitting" procedure, although the different frequencies are then fully uncoupled. We also plotted the radial and vertical profiles of the radiation fields at three wavelengths. We confirm the slower decrease at increasing altitude $z$ as compared with the radiation transfer model of Ref. \cite{Popescu-et-al2017}, indicated by the previous work \cite{A62}. Actually, when the vertical profiles of the radiation fields are calculated and plotted in a domain that involves an extrapolation to a (three times) larger domain of $z$, a slightly oscillating behaviour without a decrease at large $z$ is observed. This is explained by our study of the asymptotic behaviour of the analytical expressions of the EM field and the corresponding SED: we show that the SED calculated by the implemented model, that involves a discretization of the wave number, is a quasi-periodic function of $z$ --- although the exact SED obtained from the integral expressions (\ref{Bphi_mono})--(\ref{Ez_mono}) is order $1/z^2$ at large $z$. Thus, extrapolation on the domain of $z$ should be used parsimoniously with the current numerical implementation based on a discretization of the wave number.\\

\appendix
\section{Appendix: Discrete vs. continuous descriptions of the spectral energy density}\label{Discrete_SED}

The SED, \,$u$ \,or rather\, $u_{\bf x}$ \,(it depends on the spatial position ${\bf x}$), is normally a continuous density with respect to the wavelength or the frequency: the time-averaged volumic energy density of the EM field at some point ${\bf x}$ and in some wavelength band $[\lambda ^{(1)},\,\lambda ^{(2)}]$ is given by 
\be\label{U_1_2}
\overline{U_{1\,2}}({\bf x}) :=\overline{\frac{\delta W_{1\,2}}{\delta V}}({\bf x}) = \int _{\lambda ^{(1)}} ^{\lambda ^{(2)}} u_{\bf x}(\lambda )\, \dd \lambda .
\ee
However, in many instances, including the present work, one is led to consider a discrete spectrum, thus a finite set of frequencies, $(\omega  _j)\ (j=1,...,N_\omega )$, hence a finite set of wavelengths. It leads also to a discrete energy density, Eq. (\ref{Udiscrete}). This raises the question of how to relate together these discrete and continuous descriptions of the SED. To answer this question, we note first that the $u_j$ 's in Eq. (\ref{Udiscrete}) are indeed volumic energy densities. 
Whereas, $u_{\bf x}$ in Eq. (\ref{U_1_2}) has physical dimension $[U]/[L]$, i.e., it is 
\be\label{f_x}
f_{\bf x}(\lambda ):=\lambda  u_{\bf x}(\lambda )
\ee
which is a volumic energy density. And it is indeed $f_{\bf x}$ that is being considered by Popescu {\it et al.} \cite{Popescu-et-al2017}, when plotting the SEDs at different places in the Galaxy, or when plotting the radial or axial profiles of the radiation field at some selected wavelengths.  \\

As is more apparent with the ``separate fitting" used now (see Sect. \ref{SeparateFitting}), the discrete set of frequencies $\omega  _j$, considered in the Maxwell model of the ISRF, represents just a finite sampling of the actual continuous distribution. The link between the two descriptions is hence given simply by the following relation:
\be\label{u_j vs f}
u_j({\bf x}) = f_{\bf x}(\lambda _j) = \lambda _j u_{\bf x}(\lambda_j).
\ee
Consider a bounded spatial domain $D$. The total EM energy contained in the domain $D$ and in the wavelength band $[\lambda ^{(1)},\,\lambda ^{(2)}]$ is given, according to Eq. (\ref{U_1_2}), by
\be\label{W_1_2_D-1}
W_{1-2,\,D}:=\int_D \overline{\frac{\delta W_{1\,2}}{\delta V}} \,\dd ^3 {\bf x} = \int_D \left (\int _{\lambda ^{(1)}} ^{\lambda ^{(2)}} u_{\bf x}(\lambda )\, \dd \lambda \right ) \,\dd ^3 {\bf x} ,
\ee
i.e., using (\ref{f_x}):
\be\label{W_1_2_D-2}
W_{1-2,\,D}= \int_D \left (\int _{\lambda ^{(1)}} ^{\lambda ^{(2)}} f_{\bf x}(\lambda )\, \dd \left(\mathrm{Log}\lambda \right) \right ) \,\dd ^3 {\bf x}.
\ee
If we are using a model considering a fine-enough finite set of wavelengths $(\lambda   _j), \ j=1,...,N_\omega $, with $\lambda _1=\lambda^{(1)}$ and $\lambda _{N_\omega }=\lambda^{(2)}$, we may use (\ref{u_j vs f}) to estimate the integral over $\lambda $ in Eq. (\ref{W_1_2_D-2}) as a finite sum, e.g. a Riemann sum:
\be\label{W_1_2_D-3}
\int _{\lambda ^{(1)}} ^{\lambda ^{(2)}} f_{\bf x}(\lambda )\, \dd \left(\mathrm{Log}\lambda \right)\simeq \sum _{j=1} ^{N_\omega -1} f_{\bf x}(\lambda _j) (\mathrm{Log}\lambda _{j+1}-\mathrm{Log}\lambda _j) \simeq \sum _{j=1} ^{N_\omega -1} u_j({\bf x}) (\mathrm{Log}\lambda _{j+1}-\mathrm{Log}\lambda _j)
\ee
or a better approximation (trapezoidal, Simpson,...).


\end{document}